\documentclass[conference]{IEEEtran}

\usepackage[normalem]{ulem}

\usepackage[font={footnotesize}]{caption}
\usepackage{pgfplots,color,htor}
\pgfplotsset{compat=1.3}
\usepackage{amsfonts}
\usepackage{rotating}
\usepackage{pbox}
\usepackage{epstopdf}
\usepackage{amsmath}
\usepackage{cite}

\let\labelindent\relax
\usepackage{enumitem}

\usepackage{multirow}
\usepackage{rotating}
\usepackage{color}
\definecolor{lightgrey}{RGB}{200,200,200}
\usepackage{booktabs}

\usepackage{graphicx}
\usepackage{subfig}

\usepackage{caption}

\usepackage[absolute,overlay]{textpos}
\usepackage{listings}

\newcommand{\topologyName}{Slim Fly }

\newcommand{\htor}[1]{}
\newcommand{\mbesta}[1]{}

\newcommand{\goal}[1]{}

\usepackage{mathrsfs}

\usepackage[utf8]{inputenc}
\usepackage{cleveref}
\crefname{section}{§}{§§}
\Crefname{section}{§}{§§}

\begin{document}

\pgfplotscreateplotcyclelist{listForPlots}{%
{mark=star},
{black,mark=otimes},
{brown,mark=diamond*},
{red,mark=triangle*},
{black,mark=pentagon},
{blue,mark=+},
{blue,mark=x},
{black,mark=o},
{violet,mark=10-pointed star}}

\title{Slim Fly: A Cost Effective Low-Diameter\\ Network Topology}

\author{
\IEEEauthorblockN{Maciej Besta}
\IEEEauthorblockA{ETH Zurich\\
maciej.besta@inf.ethz.ch}
\and
\IEEEauthorblockN{Torsten Hoefler}
\IEEEauthorblockA{ETH Zurich\\
htor@inf.ethz.ch}}


\maketitle
\begin{abstract}
We introduce a high-performance cost-effective network topology called Slim Fly that
approaches the theoretically optimal network
diameter.
Slim Fly is based on graphs that approximate the solution to the degree-diameter problem.
%
We analyze Slim Fly and compare it to both traditional and state-of-the-art
networks.
Our analysis shows that Slim Fly has significant advantages over other topologies in latency, bandwidth, 
resiliency, cost, and power consumption.
%
%
Finally, we propose deadlock-free routing schemes and physical layouts for large computing centers
as well as a detailed cost and power model.
Slim Fly enables constructing cost effective and highly resilient datacenter and HPC networks
that offer low latency and high bandwidth under different HPC workloads such as stencil or graph computations.

\end{abstract}



{\vspace{0.5em}\noindent\footnotesize\textbf{This is an extended (arXiv) version of a paper published at\\ACM/IEEE Supercomputing 2014 under the same title}}

\section{Introduction}


\goal{Say interconnects and their topologies are important in HPC}


Interconnection networks play an important role in today's large-scale
computing systems. The importance of the network grows with ever
increasing per-node (multi-core) performance and memory bandwidth.
Large networks with tens of thousands of nodes are deployed in
warehouse-sized HPC and data centers~\cite{Chen:2012:LUH:2388996.2389090}. Key properties of such networks
are determined by their \emph{topologies}: the arrangement of nodes and cables.

Several metrics have to be taken into account while designing an efficient topology. First, high bandwidth is indispensable as many applications perform all-to-all communication~\cite{Tiyyagura:2008:TSP:1361718.1361719}. Second, 
networks can account for as much as 33\% of the total system cost~\cite{dally07} and 50\% of the overall system energy consumption~\cite{Abts:2010:EPD:1815961.1816004} and thus
they should be cost and power efficient. Third, low endpoint-to-endpoint latency is important for many applications, e.g., in high frequency trading. Finally, topologies should be resilient to link failures.

\goal{State why lowering network diameter can achieve above goals}

In this paper we show that \emph{lowering network diameter} not only reduces the latency but also the cost of a network and the amount of energy it consumes while maintaining high bisection bandwidth. Lowering the diameter of a network has two effects. First, it reduces energy consumption as each packet
traverses a smaller number of SerDes. Another consequence is that
packets visit fewer sinks and router buffers and will thus be less
likely to contend with other packets flowing through the network. This
enables us to reduce the number of costly routers and connections while
maintaining high bisection bandwidth. 

\goal{Show that existing topologies fail to achieve above goals}

The well-known fat tree topology~\cite{Leiserson:1985:FUN:4492.4495} is an example of a network that provides high bisection bandwidth. Still, every packet has to traverse many connections as it first has to move up the tree to reach a core router and only then go down to its destination. Other topologies, such as Dragonfly~\cite{dally08}, reduce the diameter to three, but their structure also limits bandwidth and, as we will show, has a negative effect on resiliency.



\goal{Introduce SF and state its advantages}

\begin{figure}[h!]
\vspace{-2em}
\centering
\includegraphics[width=0.46\textwidth]{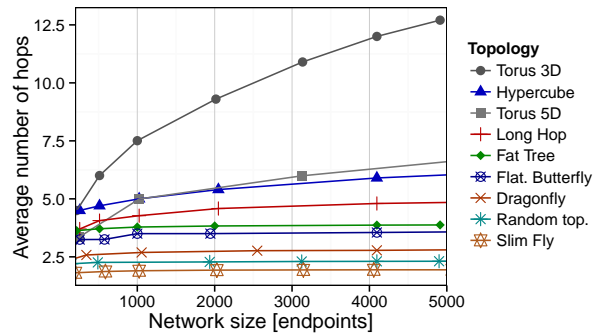}
\caption{Comparison of the average number of hops (uniform
traffic) in Slim Fly and other networks. Topologies are in balanced or close to balanced configurations (explained in Section~\ref{sec:analysis}), allowing for highest global bandwidth.}
\label{fig:hops}
\end{figure}

In this work, we propose a new topology, called Slim Fly, which 
further reduces the diameter and thus costs, energy consumption, and the latency of the
network while maintaining high bandwidth and resiliency. Slim Fly is
based on graphs with lowest diameter
for a given router radix and is, in this sense, approaching the optimal
diameter for a given router technology.
Figure~\ref{fig:hops} motivates Slim Fly by comparing the
average number of network hops for random uniform traffic
using minimal path routing on different network topologies.


\goal{Advertice SF's scalability and availability}

\footnotetext{Numbers for random topologies are updated from values obtained using the Booksim simulator to
the lower ones calculated with analytical formulas.}Slim Fly enables us to construct cost-efficient full-bandwidth networks with over 100K endpoints with diameter two using 
readily available high radix routers (e.g., 64-port Black Widow~\cite{Scott:2006:BHC:1135775.1136488} or Mellanox
108-port Director~\cite{mellanox_director}). Larger networks with up to tens of millions of endpoints can be constructed with diameter three as
discussed in Section~\ref{Slim_FlyTheory}.

\goal{State main contributions}

\noindent
The main contributions of this work are:
\begin{itemize}[leftmargin=1em]
 \item We design and analyze a new class of cost effective
   low-diameter network topologies called Slim Flies.
 \item We discuss and evaluate different deadlock-free minimal and adaptive routing
   strategies and we compare them to existing topologies and approaches.
 \item We show that, in contrast to the first intuition,
   Slim Fly, using fewer cables and routers, is more tolerant towards
   link failures than comparable Dragonflies.
 \item We show a physical layout for a datacenter or an HPC center network
   and a detailed cost and energy model.
 \item We provide a library of practical topologies with different degrees and
network sizes that can readily be used to construct
efficient Slim Fly networks\footnote{\footnotesize http://spcl.inf.ethz.ch/Research/Scalable\_Networking/SlimFly}.
The link also contains the code of all simulations from Sections~\ref{sec:analysis}-\ref{costComparison} for reproducibility.
\end{itemize}

\section{Slim Fly Topologies}

\goal{Introduce the section and describe our notation}

We now describe the main idea behind the design of Slim Fly.
Symbols used in the paper are presented in Table~\ref{tab:symbols}.

\begin{table}[h!]
\centering
\footnotesize
\begin{tabular}{r|l}
\toprule
$N$&Number of endpoints in the whole network\\
$p$&Number of endpoints attached to a router (\emph{concentration})\\
$k'$&Number of channels to other routers (\emph{network radix})\\
$k$&\emph{Router radix} ($k = k' + p$)\\
$N_r$&Number of all routers in the network\\
$D$&Network diameter\\
\bottomrule
\end{tabular}
\caption{Symbols used in the paper}
\label{tab:symbols}
\end{table}

\subsection{Construction Optimality}
\label{Slim_FlyTheory}

\goal{+Introduce MooreBound and its role in our design}

The goal of our approach is to design an \emph{optimum} or close-to-optimum 
topology that maximizes the number of endpoints $N$ for a given diameter $D$
and radix $k$ and maintains full global bandwidth. 
In order to formalize the notion of optimality we utilize the well-known concept of 
\emph{Moore Bound}~\cite{miller2005}. The Moore Bound (MB)
determines the maximum number of vertices that a potential graph with a given
$k$ and $D$ can have.
We use the MB concept in our construction scheme and we define it to be the upper limit on
the number of radix-$k$ routers that a network with a given 
diameter $D$ can contain. The Moore Bound of such a network is equal to
$N_r = 1+k' \sum_{i=0}^{D-1} (k'-1)^i$~\cite{miller2005}, where $k' = \lceil \frac{2k}{3} \rceil$
enables full global bandwidth for $D=2$ as we will show in Section~\ref{attachingEndpoints}.


\goal{+Discuss the number of endpoints in SF; lead to Degree/Diameter problem}

MB is the upper bound on the number of routers $N_r$ and thus also endpoints $N$ in the
network. For $D=2$, the maximum $N_r$
$\approx k'^2$. Thus, an example network constructed using 108-port Mellanox Director switches would have nearly
200,{}000 endpoints (we discuss the selection of the concentration $p$ in Section~\ref{attachingEndpoints}). For
$D=3$, $N_r$ is limited to $\approx k'^3$, which
would enable up to tens of millions of endpoints. Thus, we focus on graphs with diameter
two and three for relevant constructions.
To construct Slim Flies, we utilize graphs related to the well-known
\emph{degree--diameter problem}~\cite{miller2005}, which is to determine the largest graphs for a given
$k'$ and $D$.


%

\subsection{Diameter-2 Networks}
\label{diam2Construct}

\goal{+Mention the H-S graph, discuss difficulties in designing optimum graphs}

An example diameter-2 graph, which maximizes the number of vertices per given $k'$ and $D$, is the well-known Hoffman--Singleton
graph~\cite{mckay98} with 50 radix-7 vertices and 175 edges. In general, there exists no universal scheme for constructing such optimum or close-to-optimum graphs. For most $D$ and $k'$ it is not known \emph{whether} there exist optimal graphs, or \emph{how close} one can get to the Moore Bound~\cite{mckay98}.

\goal{+Introduce MMS graphs}

However, some of the introduced graphs are very close to the optimum. In order to develop a diameter-2 network we utilize a family of such graphs introduced by McKay, Miller, and {\v S}ir{\'a}n in~\cite{mckay98} (we denote them as \emph{MMS graphs}). We adopt MMS graphs and we design the Slim Fly topology (denoted as SF MMS) basing on them.
The theory of MMS graphs is deeply rooted in the \emph{graph covering techniques} and other related concepts~\cite{mckay98}. For clarity, we present a simplified construction scheme (together with an intuitive example); additional details can be found in~\cite{mckay98,siagiova01,Hafner2004223}.
%

\subsubsection{Connecting Routers}
\label{sec:conn_routers}

\goal{++Introduce the design scheme}

%
The construction of SF MMS begins with finding a prime power $q$ such that $q=4w+\delta$, where
$\delta \in \{-1,0,1\}$ and $w \in \mathbb{N}$. For such $q$
we generate an MMS graph with network radix $k'=\frac{3q-\delta}{2}$ and
number of vertices (routers) $N_r = 2 q^2$. 

\paragraph{Step 1: Constructing Base Ring $\mathbb{Z}_q$}

\goal{+++Present Step 1 (constructing the base ring)}

Let $\mathbb{Z}_q = \{0,1,...,q-1\}$ be a commutative ring with modulo addition and multiplication. We have to find a \emph{primitive element} $\xi$ of $\mathbb{Z}_q$. $\xi$ is an element of $\mathbb{Z}_q$ that \emph{generates} $\mathbb{Z}_q$: all non-zero elements of $\mathbb{Z}_q$ can be written as $\xi^{i}$ ($i \in \mathbb{N}$). In general, there exists no universal scheme for finding $\xi$~\cite{lidl1997finite}, however an exhaustive search is viable for smaller rings; all the SF MMS networks that we tested were constructed using this approach.

\paragraph{Step 2: Constructing Generator Sets $X$ and $X'$}

\goal{+++Present Step 2 (constructing generators)}

In the next step we utilize $\xi$ to construct two sets $X$ and $X'$ called \emph{generators}~\cite{Hafner2004223}. For $\delta = 1$ we have $X=\{1, \xi^2,..., \xi^{q-3}\}$ and $X'=\{\xi, \xi^3,..., \xi^{q-2}\}$ (consult~\cite{Hafner2004223} for other formulae). We will use both $X$ and $X'$ while connecting routers.

\paragraph{Step 3: Constructing and Connecting Routers}
\label{sec:constr_and_conn_routers}

\goal{+++Present step 3 (connecting routers)}

The set of all routers is a Cartesian product: $\{0,1\} \times \mathbb{Z}_q \times \mathbb{Z}_q$. Routers are connected using the following equations~\cite{Hafner2004223}:

\small
\begin{alignat}{2}
&\mbox{router } (0,x,y) \mbox{ is connected to } (0,x,y') \mbox{ iff } y-y' \in X;\\ 
&\mbox{router } (1,m,c) \mbox{ is connected to } (1,m,c') \mbox{ iff } c-c' \in X';\\ 
&\mbox{router } (0,x,y) \mbox{ is connected to } (1,m,c) \mbox{ iff } y = mx + c;
\end{alignat}
\normalsize

%

\goal{+++Present the design intuition}

\begin{figure}[h!]
\centering
\includegraphics[width=0.47\textwidth]{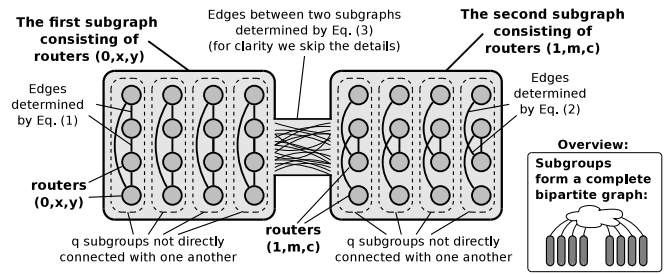}
\caption{General structure of the MMS graph (\cref{sec:conn_routers}).
}
\label{fig:mms_general}
\end{figure}

Intuitively, MMS graphs have highly symmetric internal structure: they consist of two subgraphs, each composed of the \emph{same} number of \emph{identical} subgroups of routers. The first subgraph is composed of routers $(0,x,y)$ while the other consists of routers $(1,m,c)$. An overview is presented in Figure~\ref{fig:mms_general}. We will use this property while designing a physical layout for a datacenter or an HPC center in Section~\ref{arrangementInCabinets}.


\begin{figure*}
\centering
\begin{tabular}{cc}
 \subfloat[Connections between routers in each subgraph (\cref{sec:conn_routers}, Eq. (1)-(2)). Note that respective groups have identical connection patterns.]{
  \includegraphics[width=0.41\textwidth]{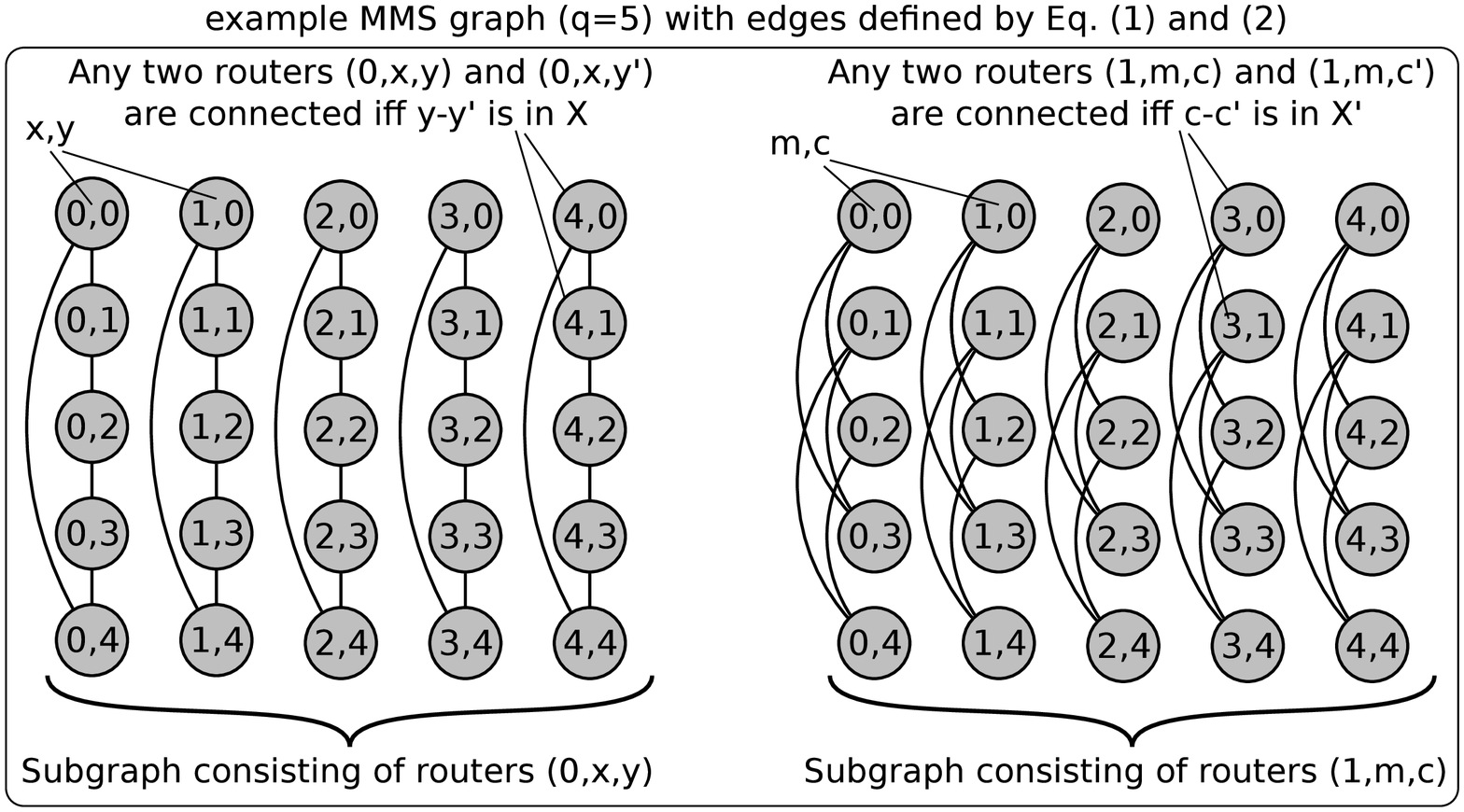}
  \label{fig:mms_inner}
 }
 &
 \subfloat[Connections between two subgraphs (\cref{sec:conn_routers}, Eq. (3)). For clarity we present only the edges originating at $(1,0,0)$ and $(1,1,0)$.]{
  \includegraphics[width=0.41\textwidth]{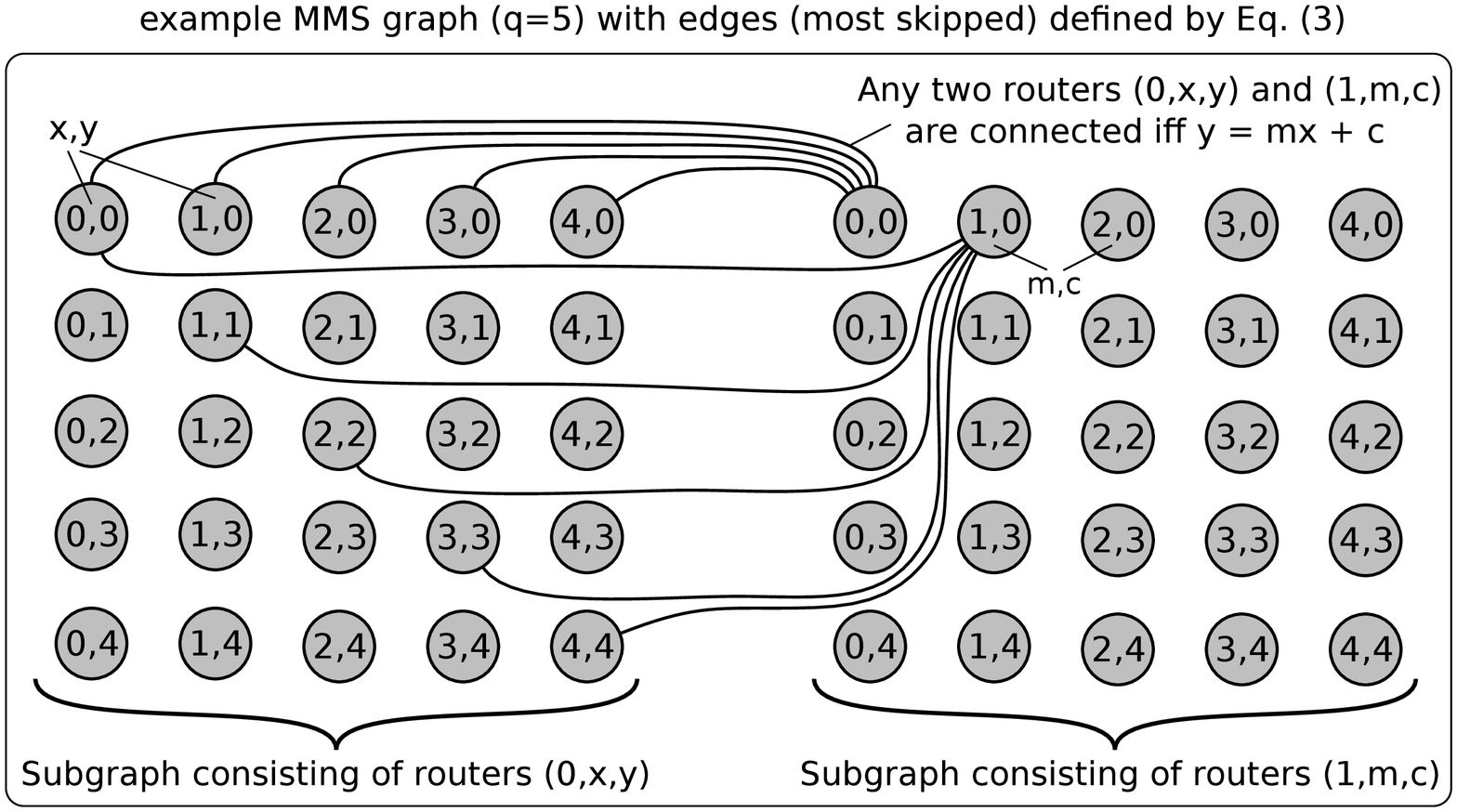}
  \label{fig:mms_outer}
 }
 \end{tabular}
\caption{Connecting routers in an MMS graph ($q=5$). For clarity, we denote routers $(0,x,y)$ as $x,y$; and routers $(1,m,c)$ as $m,c$.}
\label{fig:mms_example}
\end{figure*}

\paragraph{Example MMS Construction for $q=5$}
\label{sec:example_mms_g_5}

\goal{+++Present steps 1-2 for the example H-S construction}

We now construct an example MMS (the Hoffman-Singleton graph) to illustrate the presented scheme in practice. We select $q=5$, thus $\mathbb{Z}_5 = \{0,1,2,3,4\}$ and the primitive element $\xi = 2$. We can verify it easily by checking that: $1 = \xi^4\;mod\;5 = 2^4\;mod\;5$, $2 = 2^1\;mod\;5$, $3 = 2^3\;mod\;5$, $4 = 2^2\;mod\;5$.
The construction of generator sets is also straightforward: $X = \{1, 2^2, ..., 2^{5-3}\} = \{1,4\}$ and $X' = \{2, 2^3, ..., 2^{5-2}\} = \{2,3\}$ (all operations are of course done modulo $q$).

\goal{+++Present step 3 for the example H-S construction}

The router set of our SF MMS is $\{0,1\} \times \mathbb{Z}_5 \times \mathbb{Z}_5$. We apply Equation~(1) to connect routers $(0,x,y)$. Then, we use  Equation~(2) for routers $(1,m,c)$. The results are shown in Figure~\ref{fig:mms_inner}; for clarity, we denote routers $(0,x,y)$ as $x,y$; and routers $(1,m,c)$ as $m,c$. Finally, we apply Equation~(3) to connect routers $(0,x,y)$ with $(1,m,c)$ (see Figure~\ref{fig:mms_outer}).

\subsubsection{Attaching Endpoints}
\label{attachingEndpoints}

\goal{++Derive the concentration for full global bandwidth (Part 1)}

We now illustrate our formula for $p$ (concentration) that ensures full global bandwidth.
%
%
The global bandwidth of a network is defined as the theoretical 
cumulative throughput if all processes simultaneously communicate with
all other processes in a steady state.
To maximize the global bandwidth of SF MMS, we first consider the network 
channel load (we model each full-duplex link with two channels, one in
each direction): each router can reach $k'$ routers in distance one and
$N_r-k'-1$ routers in distance two. The whole network has a total number of
$k'\cdot N_r$ channels.
We define the channel load $l$ as the average number of routes (assuming
minimal routing) that lead through each link of the network.
We have $p$ endpoints per router and each router forwards messages
to approximately $p\cdot N_r$ destinations from each local endpoint. We
get a total average load per channel $l=\frac{(k'+2\cdot(N_r-k'-1))\cdot
p^2N_r}{k'N_r}=\frac{(2N_r-k'-2)\cdot p^2}{k'}$.

\goal{++Derive the concentration for full global bandwidth (Part 2)}

\begin{figure}[h!]
\centering
\includegraphics[width=0.45\textwidth]{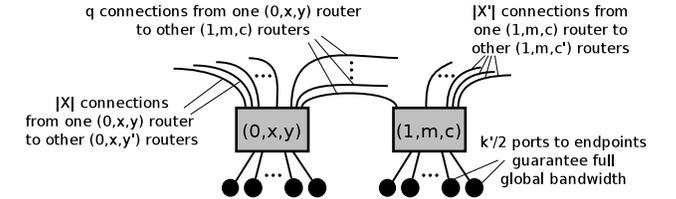}
\caption{Connecting routers and endpoints in SF MMS.}
\label{fig:mms_endpoints}
\end{figure}

Each endpoint injects to approximately $N=pN_r$ destinations through its
single uplink. 
A network is called \emph{balanced} if each endpoint can inject at full
capacity, i.e., $pN_r = [(2N_r-k'-2)\cdot p^2]/k'$. Thus, we
pick the number of endpoints per router $p \approx
\frac{k'N_r}{2N_r-k'-2}=\frac{N_r}{l}$ to achieve full global bandwidth.
Finally, we get $p\approx k'/(2-\frac{k'}{N_r}-\frac{2}{N_r}) \approx
\left\lceil k'/2\right\rceil$ which means that $\approx67\%$ of each router's ports
connect to the network and $\approx33\%$ of the ports connect to
endpoints.
%
%
%
%
An overview of the connections originating at a single router is presented in Figure~\ref{fig:mms_endpoints}.

\begin{figure*}
\centering
 \subfloat[The MB and graphs with $D=2$ (\cref{sec:mb_diam_2_compare}).]{
  \includegraphics[width=0.305\textwidth]{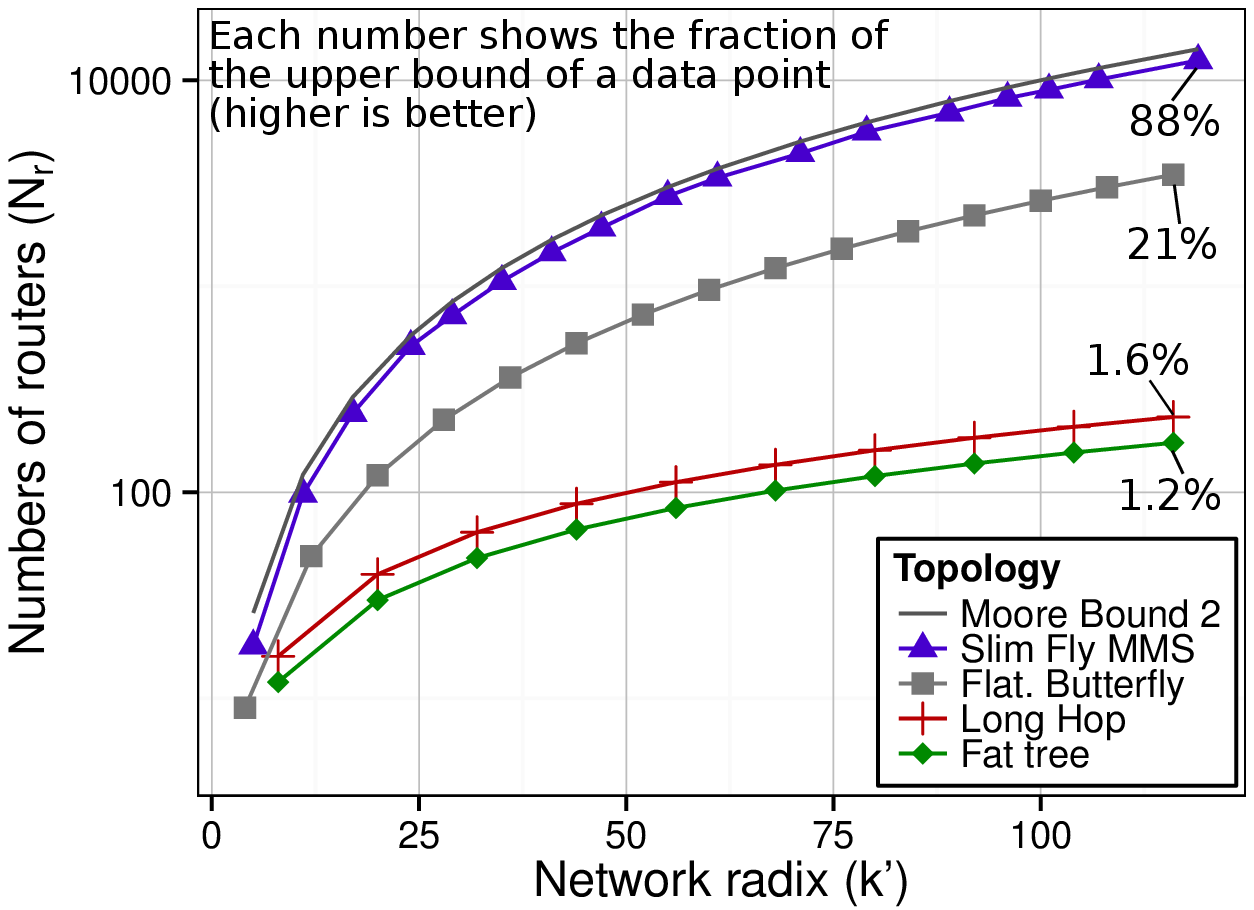}
  \label{fig:MB}
 }\hfill
 \subfloat[The MB and graphs with $D=3$ (\cref{sec:mb_diam_3_compare}).]{
  \includegraphics[width=0.315\textwidth]{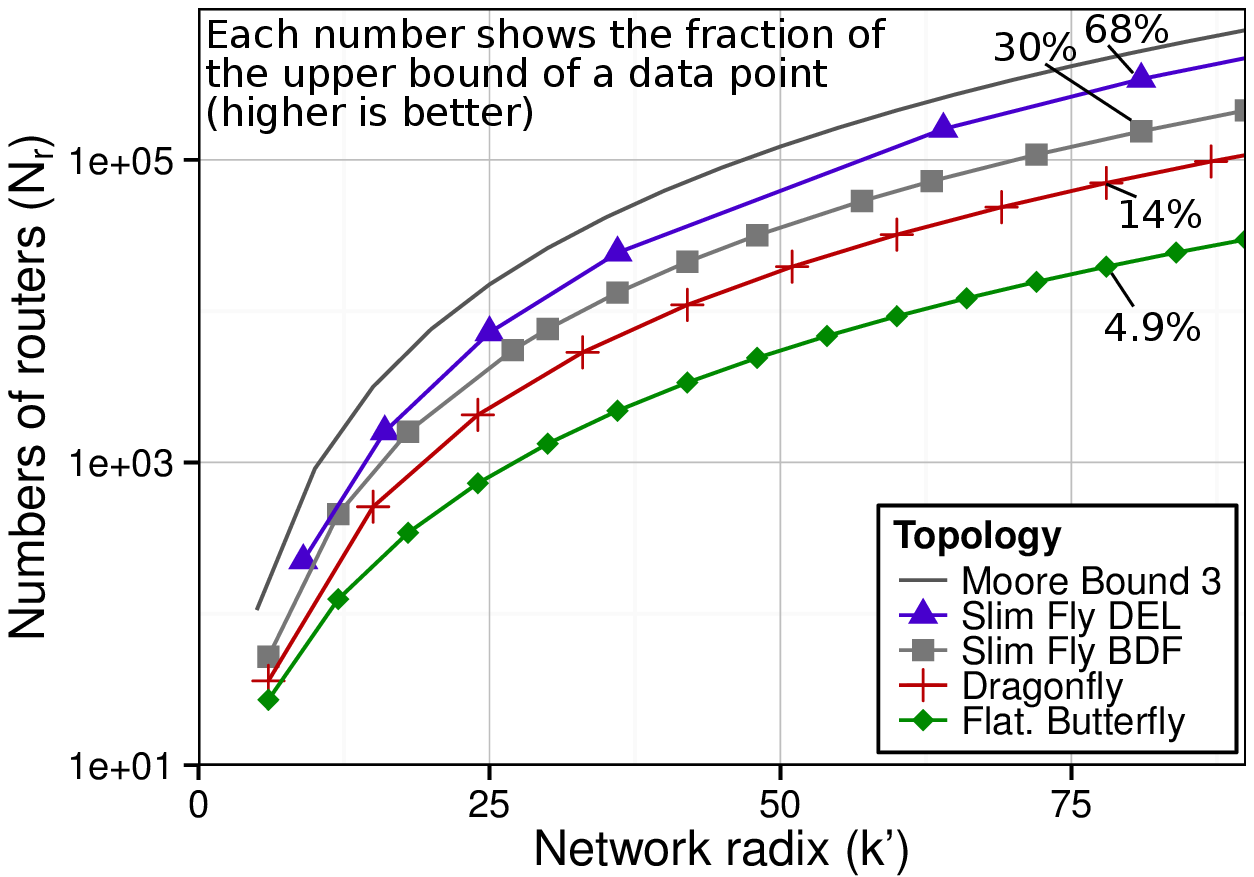}
  \label{fig:MB3}
 }
 \hfill 
 \subfloat[Bisection bandwidth (BB) comparison (\cref{sec:bb}).]{
  \includegraphics[width=0.33\textwidth]{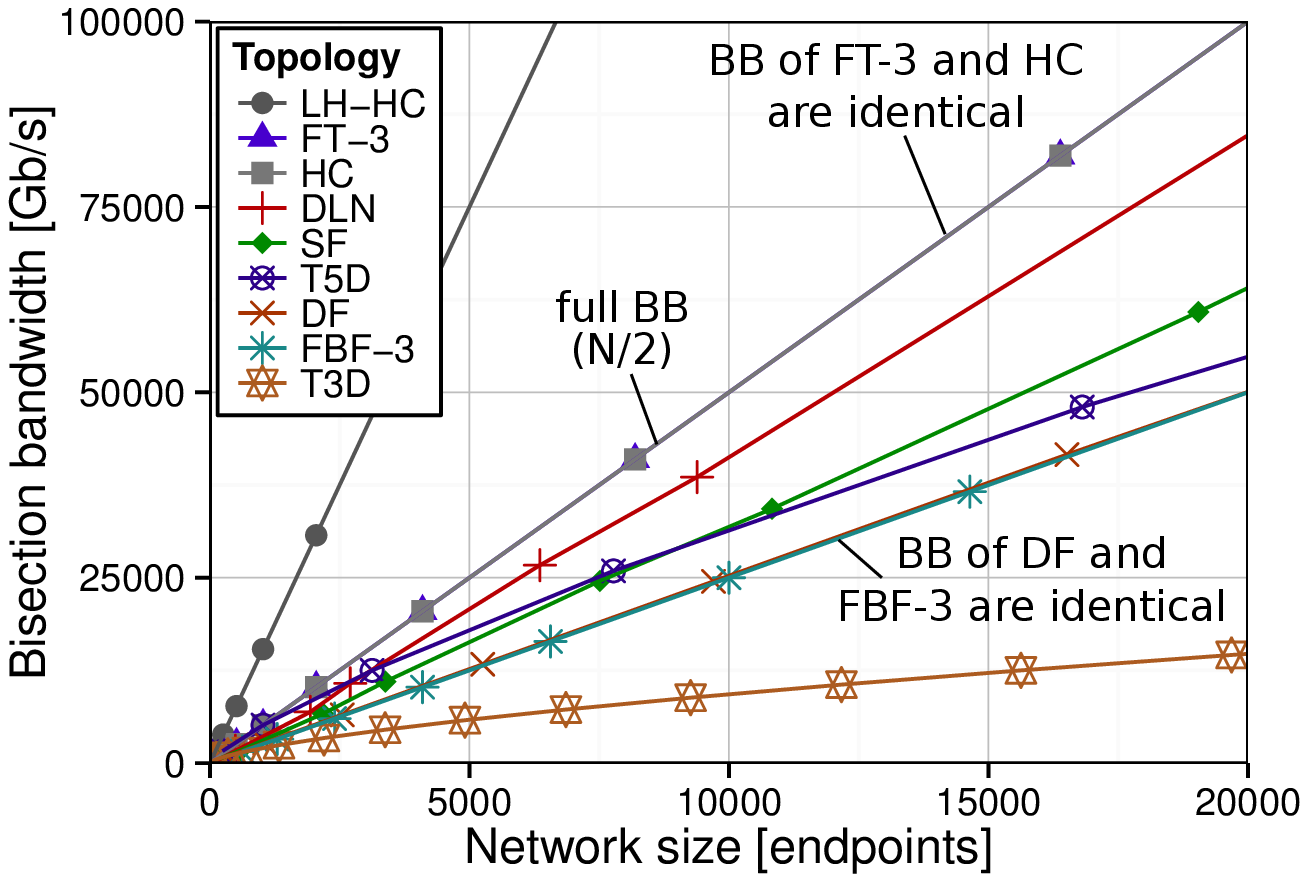}
  \label{fig:BB}
 }
 \caption{Comparison of the Moore Bound (for diameter 2 and 3 constructions) and bisection bandwidth (we assume 10 Gb/s/link). For the Moore Bound comparison we skip networks which does not have constant diameter (random topologies, hypercube, and tori).}
\label{fig:mb_bb}
\end{figure*}

\subsubsection{Comparison to Optimality (the Moore Bound)}
\label{sec:mb_diam_2_compare}

\goal{++Show SF MMS is close-to-optimum and much better than others}

Figure~\ref{fig:MB} compares the distance between topologies with $D=2$ and the MB. We
see that SF MMS is very close to the optimum. For $k' = 96$, MMS has 8,{}192 routers, which is only 12\% worse than the upper bound (9,{}217). Other topologies (a Long Hop described in Section E-S-1 of~\cite{2013arXiv1301.4177T}, a two-stage fat tree, and a two-level Flattened Butterfly) are up to several orders of magnitude worse. Thus, in the paper we do not compare to these topologies, as they cannot be easily used to construct networks of practical size (e.g., a Long Hop with merely 50,{}000 endpoints requires routers with radix $\approx$340).

\subsection{Diameter 3 Networks}
\label{sec:mb_diam_3_compare}

\goal{+Present diameter 3 graphs (BDF/DEL) that approach MB}


We present two classes of graphs that approach the
MB for $D=3$. \emph{Bermond, Delorme and Fahri (BDF) graphs} can be generated using a scheme described in~\cite{bermond82}.
They have $k' = \frac{3(u+1)}{2}$ and $N_r = \frac{8}{27}k'^3 -
  \frac{4}{9}k'^2 + \frac{2}{3}k'$ for a given odd prime power $u$.
The second class are \emph{Delorme (DEL) graphs}~\cite{delorme85} characterized by
$N_r = (v+1)^2 (v^2+1)^2$ and $k' = (v+1)^2$ for
  a given prime power $v$.
  
\goal{+Show SF BDF/DEL is close-to-optimum and much better than others}

Figure~\ref{fig:MB3} compares the number of routers in BDF and DEL
graphs with two other networks that have $D=3$: Dragonfly and 3-level Flattened Butterfly. Dragonfly achieves only 14\% (e.g. for $k'=96$) of the maximum possible number of routers for a given $k'$ and $D=3$; Flattened Butterfly is $\approx$3 times worse. Delorme and BDF graphs achieve, respectively, 68\% and 30\% of the Moore Bound.
  
\goal{+Explain why we limit the analysis to SF MMS}  
  

\subsubsection{Constructing BDF graphs}

\paragraph{The $*$ Product}

Consider two graphs $G_1=(V_1,E_1)$ and $G_2=(V_2,E_2)$.
Let $U$ be a set of arcs constructed from the edges of $G_1$ by
taking an arbitrary orientation. Let $f_{(x,y)}$ be a one-to-one
mapping from $V_2$ to itself, for any arc $(x,y)$~\cite{bermond82}.

The $*$ product creates a new graph $G' = (V',E') = G_1 * G_2$.
$V'$ is a cartesian product $V_1 \times V_2$. A vertex $(a_1,a_2) \in V'$ 
is connected to $(b_1,b_2) \in V'$ iff either $a_1 = b_1 \land \{a_2,b_2\} \in E_2$,
or $(a_1,b_1) \in U \land b_2 = f_{(a_1,b_1)}(a_2)$~\cite{bermond82}.

\paragraph{Constructing Graph $P_u$}

Assume $u$ is an odd prime power. We know that there exists a projective plane of order $u$
with a numbering of the points ($M_i$) and of the lines ($D_j$)~\cite{bermond82} where
$1 \geq M_i,D_j \leq u^2 + u + 1$ and $M_i \in D_j \Rightarrow M_j \in D_i$.
We now construct a graph $P_u$ from the points of the projective plane. Two vertices
$M_i$ and $M_j$ are connected iff $M_j \in D_i$. $P_u$ has a diameter of 2, degree $u+1$,
and $u^2+u+1$ vertices~\cite{bermond82}.

%

\paragraph{Constructing the Final Graph $P_u * G_{k'/3}$}

We say that a graph $G=(V,E)$ has property $P^{*}$ iff it has a diameter of at most 2
and there exists an involution $f$ of $V$ such that $\forall_{v \in V}:\quad V = \{v\} \cup \{f(v)\} \cup \{f(\Gamma(v))\} \cup \{\Gamma(f(v))\}$~\cite{bermond82}.

Let $k' = \frac{3(u+1)}{2}$ and let $G_{k'/3}$ be a graph satisfying property $P^{*}$
(see~\cite{bermond82} for construction details). A graph $P_u * G_{k'/3}$ has
diameter 3, degree $k'$, and $N_r = \frac{8}{27}k'^3 -
  \frac{4}{9}k'^2 + \frac{2}{3}k'$.

In this work, we focus on MMS graphs because their scalability suffices
for most large-scale networks having more than 100K endpoints. Analyses with the diameter three
constructions show lower but similar results in terms of cost and
performance benefits over other topologies since they approach the optimal structure.

\section{Slim Fly Structure Analysis}
\label{sec:analysis}

\goal{Introduce the section and discuss compared topologies}

We now analyze the structure of SF MMS in terms of common metrics: network diameter,
average distance, bisection bandwidth, and resiliency.
%
We compare Slim Fly to the topologies presented in Table~\ref{tab:compar}.
Most of them are established and well-known designs and we refer the reader
to given references for more details.
%
\verb$DLN$\footnote{We use random topologies that are generated basing on a ring. Koibuchi et al. denote them as \texttt{DLN-2-y}, where \texttt{y} is the number of additional random shortcuts added to each vertex~\cite{DBLP:conf/isca/KoibuchiMAHC12}.} are constructed from a ring topology by adding random edges identified by a number of routers and degree~\cite{DBLP:conf/isca/KoibuchiMAHC12}. Long Hops are networks constructed from Cayley graphs using optimal error correcting codes~\cite{2013arXiv1301.4177T}
We utilize a variant of Long Hops that augments hypercubes (introduced in Section E-S-3 of~\cite{2013arXiv1301.4177T}).


\textbf{Topology parameters}
For high radix networks
we select the concentration $p$ to enable {balanced} topology variants
with full global bandwidth. Respective values of $p$, expressed as a function
of radix $k$, are as follows:
$p = \lfloor (k+1)/4 \rfloor$
(\verb$DF$), $p = \lfloor (k+3)/4 \rfloor$ (\verb$FBF-3$),
$p = \lfloor \sqrt{k} \rfloor$ (\verb$DLN$), $p = \lfloor k/2 \rfloor$ (\verb$FT-3$).
For lower radix topologies (\verb$T3D$, \verb$T5D$, \verb$HC$, \verb$LH-HC$)
we select $p = 1$ following strategies from~\cite{abts2011cray,dally07,Kim:2007:FBT:1331699.1331717}.
%


\subsection{Network Diameter}
\label{sec:diam_analysis}


\goal{+Discuss network diameter}

The structure of MMS graphs ensures that
\verb$SF$'s diameter is 2. The comparison to other topologies is illustrated in Table~\ref{tab:compar}. For \texttt{LH-HC} we report the values for generated topologies of size from $2^{8}$ to $2^{13}$ endpoints ($D$ increases as we add endpoints).
Numbers for \texttt{DLN} come from~\cite{DBLP:conf/isca/KoibuchiMAHC12}. \verb$SF$ offers the lowest diameter out of all compared topologies.

\begin{table}[h!]
\footnotesize
\centering
\begin{tabular}{l|l|l|l}
\toprule
Topology&Symbol&Example System&Diameter\\
\midrule
3-dimensional torus~\cite{Alverson:2010:GSI:1901617.1902283}&\verb$T3D$& Cray Gemini~\cite{Alverson:2010:GSI:1901617.1902283} &$\lceil 3/2 \sqrt[3]{N_r}\rceil$\\
5-dimensional torus~\cite{Chen:2011:IBG:2063384.2063419}&\verb$T5D$& IBM BlueGene/Q~\cite{Chen:2012:LUH:2388996.2389090}&$\lceil 5/2 \sqrt[5]{N_r}\rceil$\\
Hypercube~\cite{pleiades}&\verb$HC$& NASA Pleiades~\cite{pleiades} &$\lceil\log_2N_r\rceil$\\
3-level fat tree~\cite{Leiserson:1985:FUN:4492.4495}&\verb$FT-3$& Tianhe-2~\cite{dongarra2013visit} &4\\
3-level Flat. Butterfly~\cite{dally07}&\verb$FBF-3$& - &3\\
Dragonfly topologies~\cite{dally08}&\verb$DF$& Cray Cascade~\cite{DBLP:conf/sc/FaanesBRCFAJKHR12} &3\\
Random topologies~\cite{DBLP:conf/isca/KoibuchiMAHC12}&\verb$DLN$& - &3--10\\
Long Hop topologies~\cite{2013arXiv1301.4177T}&\verb$LH-HC$& Infinetics Systems~\cite{2013arXiv1301.4177T} &4--6\\
\midrule
\textbf{Slim Fly MMS}&\textbf{\ttfamily{SF}}& - &\textbf{2}\\
\bottomrule
\end{tabular}
\caption{Topologies compared in the paper, their diameters (\cref{sec:diam_analysis}), and example existing HPC systems that use respective topologies.}
\label{tab:compar}
\end{table}


\subsection{Average distance}

\goal{+Discuss network average distance}

The distance between any two endpoints in \verb$SF$ is
always equal to or smaller than two hops. We compare \verb$SF$ to
other topologies in Figure~\ref{fig:hops}. The average distance
is asymptotically approaching the network diameter for all
considered topologies and is lowest for \verb$SF$ for all analyzed network sizes.

\subsection{Bisection Bandwidth}
\label{sec:bb}

\goal{+Discuss bisection bandwidth}

Figure~\ref{fig:BB}
presents the bisection bandwidth (BB) of compared topologies. For \verb$SF$ and \verb$DLN$ we approximate the bisection bandwidth using the 
METIS~\cite{karypis99} partitioner.
Bisection bandwidths for other topologies can be derived
analytically and are equal to: $\lfloor \frac{N}{2} \rfloor$ (\verb$HC$ and \verb$FT-3$), 
$\lfloor \frac{2N}{k'} \rfloor$ (tori), and $\lfloor \frac{N+2p^2-1}{4} \rfloor \approx \lfloor \frac{N}{4} \rfloor$ (\verb$DF$ and \verb$FBF-3$)
~\cite{dally08,dally07,Leiserson:1985:FUN:4492.4495,Dally90performanceanalysis,2013arXiv1301.4177T}.
%
\verb$LH-HC$ has the bandwidth of $\lfloor \frac{3 N}{2} \rfloor$ as it was designed specifically to increase bisection bandwidth. \verb$SF$ offers higher bandwidth than
\verb$DF$, \verb$FBF-3$, \verb$T3D$, and \verb$T5D$.

\subsection{Resiliency}

\goal{+Introduce the section}

We compare \verb$SF$ to other topologies using three different resiliency metrics. To prevent deadlocks
in case of link failures, one may
utilize Deadlock-Free Single Source Shortest Path (DFSSSP) routing~\cite{domke-hoefler-dfsssp} (see Section~\ref{sec:df_df} for details).

\subsubsection{Disconnection Metrics}
\label{sec:disc_analysis}

\goal{++Discuss the disconnection metrics, the methodology, and the results}

We first study how many random links have to be removed before a network becomes disconnected. We simulate random
failures of cables in 5\% increments with enough samples to guarantee a
95\% confidence interval of width 2. Table~\ref{tab:resiliency-discdd} illustrates the results of the analysis.
%
The three most resilient topologies are \verb$SF$, \verb$DLN$, and \verb$FBF-3$.
Interestingly, random topologies, all with
diameter three in our examples, are very resilient, and one can
remove up to 75\% of the links before the network is disconnected. This can be explained with the emergence of the \emph{giant
component} known from random graph theory~\cite{Bollobas01:random_graphs}.
\verb$FBF-3$ is also resilient thanks to high path diversity.
\verb$DF$, also diameter three,
is less resilient due to its structure, where a failure in a global
link can be disruptive. A similar argumentation applies to \verb$FT-3$.
For torus networks, the resilience level decreases as we increase $N$. This is due to a fixed radix that
makes it easier to disconnect bigger networks. Finally, the resilience level of both \verb$HC$ and \verb$LH-HC$ 
does not change with $N$. The radix of both networks increases together with $N$, which prevents the resilience level
from decreasing as in tori. Still, the rate of this increase is too slow to enable gains in resilience as in high radix topologies.

\begin{table}[h!]
\centering
\footnotesize
\renewcommand{\tabcolsep}{1.4mm}
\begin{tabular}{l|ccccccccc}

\toprule
$\approx N$ & \verb$T3D$ & \verb$T5D$ & \verb$HC$ & \verb$LH-HC$ & \verb$FT-3$ & \verb$DF$ & \verb$FBF-3$ & \verb$DLN$ & \textbf{\ttfamily{SF}} \\

\midrule
256 & 25\% & 50\% & 40\% & 55\% & - & 45\% & 50\% & - & \textbf{45\%} \\
512 & - & - & 40\% & 55\% & 35\% & - & 55\% & 60\% & \textbf{60\%} \\
1024 & 15\% & 40\% & 40\% & 55\% & 40\% & 50\% & 60\% & - & \textbf{-} \\
2048 & 10\% & - & 40\% & 55\% & 40\% & 55\% & 65\% & 65\% & \textbf{65\%} \\
4096 & 5\% & 40\% & 45\% & 55\% & 55\% & 60\% & 70\% & 70\% & \textbf{70\%} \\
8192 & 5\% & 35\% & 45\% & 55\% & 60\% & 65\% & - & 75\% & \textbf{75\%} \\

\bottomrule
\end{tabular}
\caption{Disconnection Resiliency (\cref{sec:disc_analysis}): the maximum
number of cables that can be removed before the network is disconnected.
Missing values indicate the inadequacy of a balanced topology variant for a given $N$.}
\label{tab:resiliency-discdd}
\end{table}


\goal{++Discuss the results for SF}

\verb$SF$, the only topology
with $D=2$, is highly resilient as its structure provides high path diversity. As we will show in Section~\ref{arrangementInCabinets}, \verb$SF$ has a modular layout similar to \verb$DF$. However, instead of one link between groups of routers there are $2q$ such links, which dampens the results of a global link failure.

\subsubsection{Increase in Diameter}

\goal{++Discuss the increase in diameter metrics and the results}

Similarly to Koibuchi et al.~\cite{DBLP:conf/isca/KoibuchiMAHC12}, we also characterize
the resiliency by the increase in diameter while removing links
randomly. For our analysis, we make the (arbitrary) assumption that an increase
of up to two in diameter can be tolerated. The relative results are similar to the ones obtained for disconnection metrics. The only major difference is that non-constant diameter topologies such as tori are now rather resilient to faults because random failures are unlikely to lie on a critical path. For a network size $N=2^{13}$, \verb$SF$ can withstand up to 40\% link failures before the diameter grows beyond
four. The resilience of \verb$SF$ is slightly worse than \verb$DLN$ (tolerates up to 60\% link failures), comparable to tori, and significantly better than \verb$DF$ (withstands 25\% link failures).

\subsubsection{Increase in Average Path Length}

\goal{++Discuss the increase in avg path metrics and the results}

While the diameter may be important for certain latency-critical
applications, other applications benefit from a short average path length
(which may also increase the effective global bandwidth). Thus, we also
investigate the resiliency of the average path length of the topologies.
We assume that an increase of one hop in the average distance between
two nodes can be tolerated. Again, this is an arbitrary value for the
purpose of comparison. The results follow a similar pattern as for the diameter metrics. Tori
survive up to 55\% link failures. \verb$DLN$ is most resilient and
can sustain up to 60\% link failures for a network with $N=2^{13}$.
\verb$DF$ withstands up to 45\% of link crashes.
\verb$SF$ is again highly resilient and it
tolerates up to 55\% link failures.


%
%
%

\section{Routing}
\label{routing}

\goal{Introduce the section and the notation}

We now discuss minimal and non--minimal routing
for \verb$SF$ and we present a UGAL--L
(global adaptive routing using local information) algorithm suited for \verb$SF$ together
with the comparison to UGAL--G
as defined in~\cite{ugal2005scheme}. We also show how to guarantee deadlock-freedom in \verb$SF$.
We consider routing packets from a source endpoint $s$ attached to
a router $R_s$ to a destination endpoint $d$ connected to a router $R_d$.

\subsection{Minimal Static Routing}


\goal{+Explain and discuss MIN routing}

In minimal (MIN) routing in \verb$SF$ a packet
is routed either directly (if $R_s$ is connected to $R_d$) or using two hops if the distance between $R_s$ and $R_d$ is two.
Such minimal routing can easily be implemented with current statically
routed networking technologies such as InfiniBand or Ethernet.

\subsection{Valiant Random Routing}

\goal{+Explain and discuss VAL routing}

The Valiant Random Routing (VAL) algorithm \cite{valiant1982scheme} can
be used for \topologyName to load--balance adversarial traffic
scenarios for which minimum routing is inefficient. To route a packet,
the protocol first randomly selects a router $R_r$ different from $R_s$ and $R_d$.
The packet is then routed along two minimal paths: from $R_s$ to $R_r$, and from $R_r$ to $R_d$.
%
%
%
%
Paths generated by VAL may consist of 2, 3, or 4 hops, depending on whether routers $R_s$, $R_r$, and $R_d$ are directly connected. One may also impose a constraint on a selected random path so that it contains at most 3 hops. However, our simulations indicate that this results in higher average packet latency because it limits the number of available paths (we discuss our simulation infrastructure and methodology in detail in Section~\ref{performance}).

\mbesta{add a section in simulation that discusses briefly results for the protocol variants}


\subsection{Non--minimal Adaptive Routing}

\goal{+Introduce UGAL}

The Universal Globally--Adaptive Load--balanced (UGAL) algorithm
\cite{ugal2005scheme} selects either a minimum or a VAL--generated path for
a packet basing on hop distance and sizes of queues between two endpoints.
For \verb$SF$ we investigate two variants.

\subsubsection{Global UGAL Version (UGAL--G)}

\goal{++Discuss UGAL-G}

UGAL--G has access to the sizes of all router queues in
the network. For each injected packet it generates a set of random VAL
paths, compares them with the MIN path, and selects a path with
the smallest sum of output router queues. Our simulations indicate that the choice of
4 paths provides the best average packet latency. UGAL--G approximates the
\emph{ideal} implementation of UGAL routing and thus provides a
good way to evaluate the quality of the local version.

\subsubsection{Local UGAL Version (UGAL--L)}

\goal{++Discuss UGAL-L}


UGAL--L can only access the local output queues at
each router. To route a packet, it first generates a set of VAL paths and computes the MIN path.
Then, it multiplies the length of each path (in hops) by the
local output queue length, and
picks the one with the lowest result.
The number of generated random paths influences the simulation results. We compared implementations using between 2 and 10 random selections and we find empirically that selecting 4 results in lower overall latency.


\begin{figure*}
\centering
 \subfloat[Random traffic (\cref{uniformTraffic}).]{
  \includegraphics[width=0.235\textwidth]{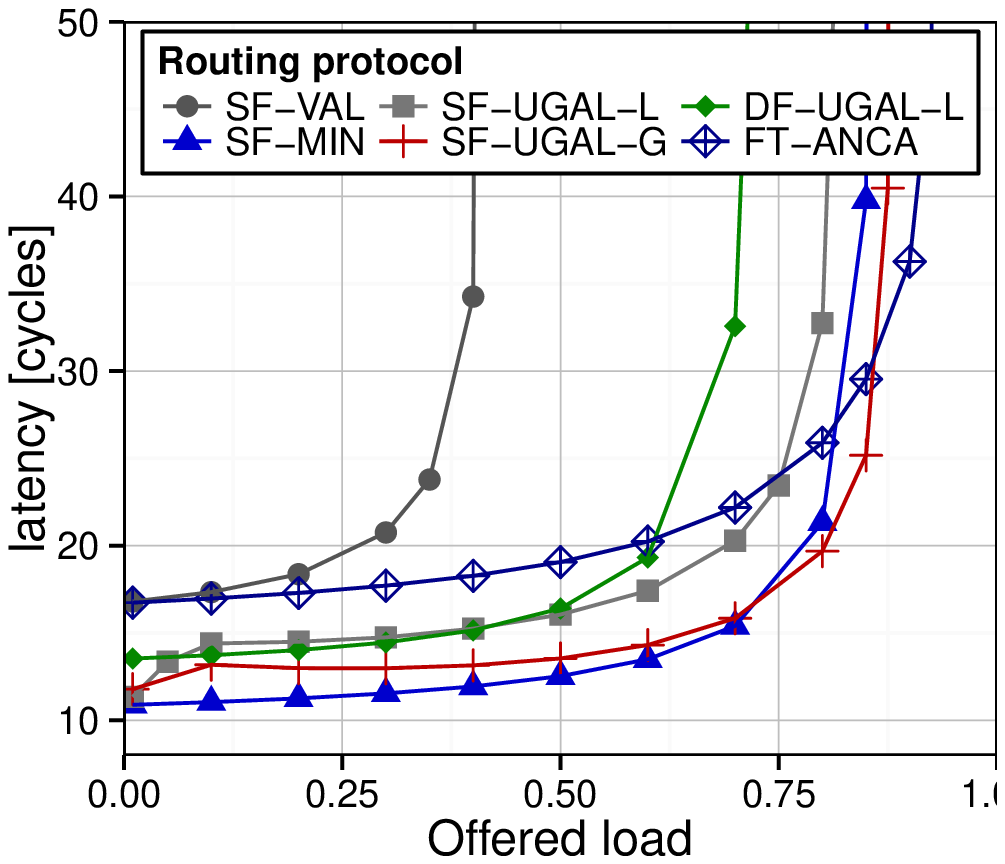}
  \label{fig:simSlimUniformTraffic64}
 }\hfill
 \subfloat[Bit reverse traffic (\cref{shuffle}).]{
  \includegraphics[width=0.235\textwidth]{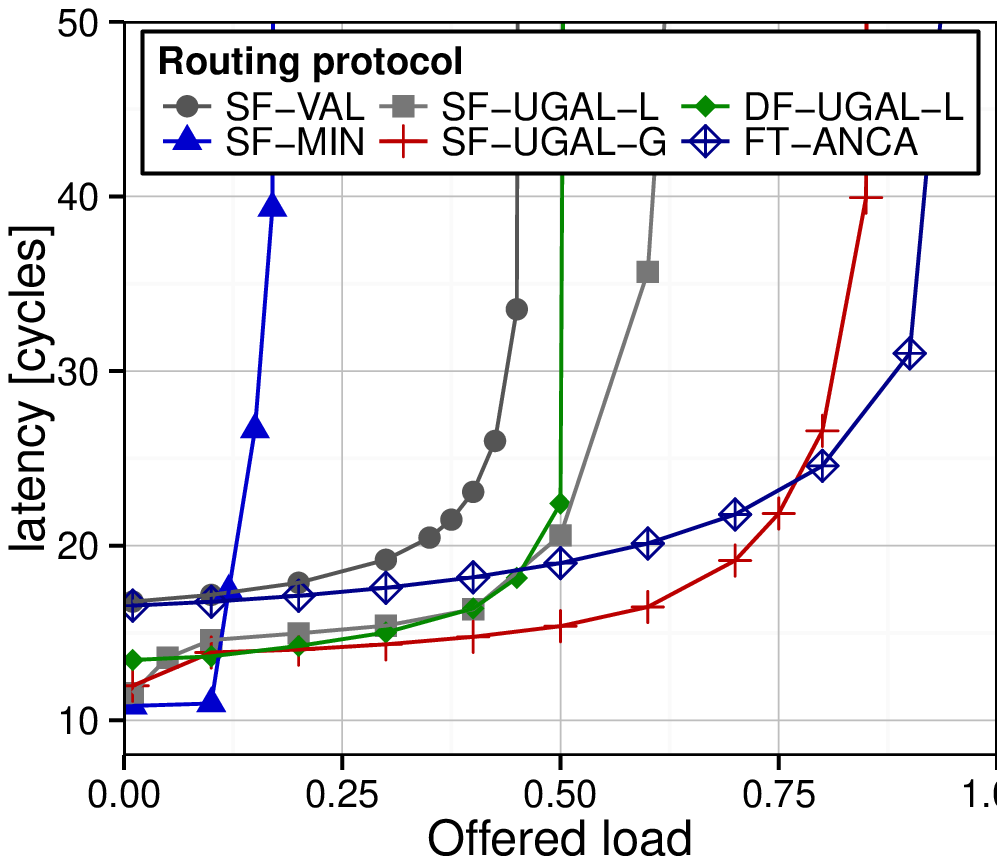}
  \label{fig:simSlimBitrevTraffic64}
 }
   \hfill
 \subfloat[Shift traffic (\cref{shuffle}).]{
  \includegraphics[width=0.235\textwidth]{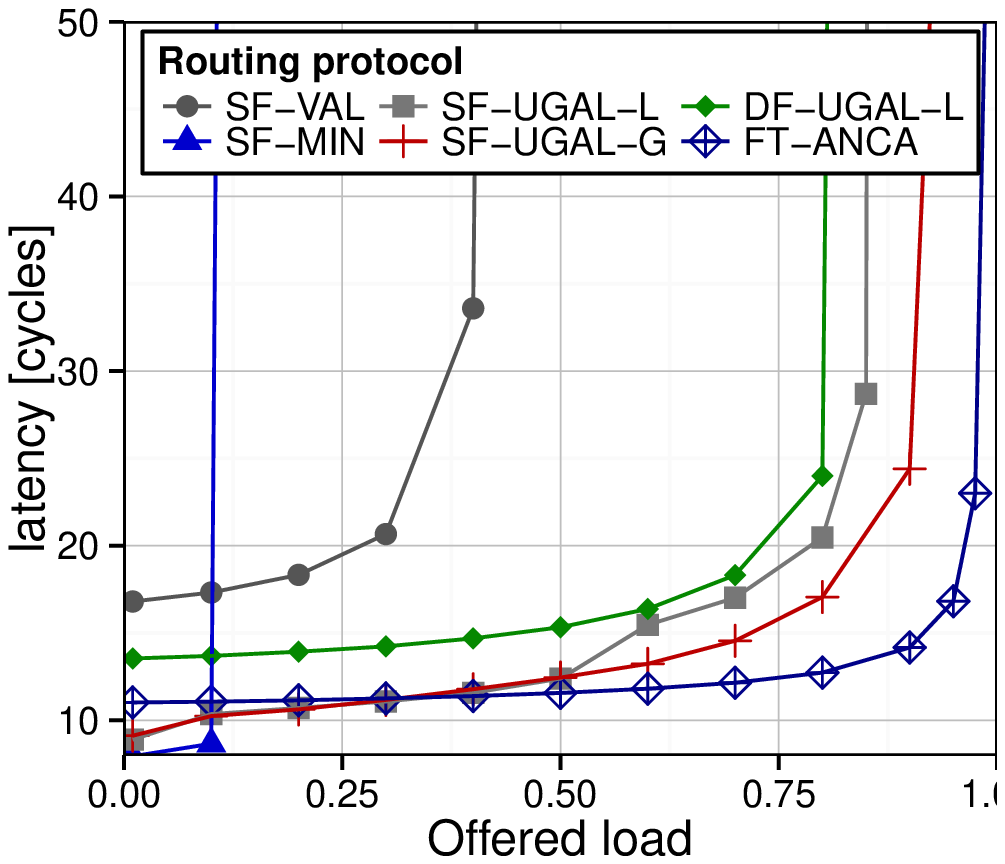}
  \label{fig:simSlimAsymTraffic64}
 }
   \hfill 
 \subfloat[Worst-case traffic (\cref{WCScenario}).]{
  \includegraphics[width=0.235\textwidth]{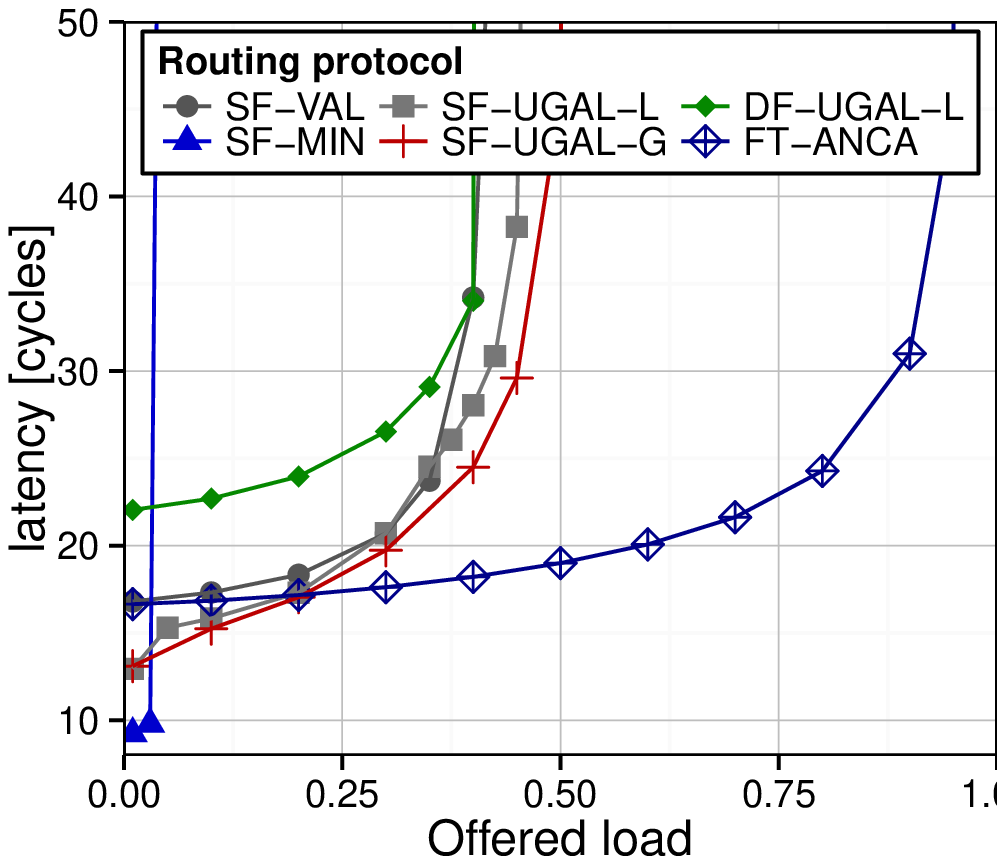}
  \label{fig:simSlimWCTraffic64}
 }
 \caption{Performance comparison of \texttt{SF}, \texttt{DF}, and \texttt{FT-3}. We use different protocols in the Slim Fly analysis: minimum static routing (\texttt{SF-MIN}), Valiant (\texttt{SF-VAL}), UGAL-L (\texttt{SF-UGAL-L}), UGAL-G (\texttt{SF-UGAL-G}). For \texttt{DF} and \texttt{FT-3} we use Dragonfly UGAL-L (\texttt{DF-UGAL-L}) and Adaptive Nearest Common Ancestor (\texttt{FT-ANCA}), respectively. We use the buffer size of 64 flit entries.}
\label{fig:perf}
\end{figure*}

\subsection{Deadlock-Freedom}
\label{sec:df_df}

\goal{+Describe two general strategies for deadlock-freedom}

\begin{figure}[h!]
\centering
\includegraphics[width=0.4\textwidth]{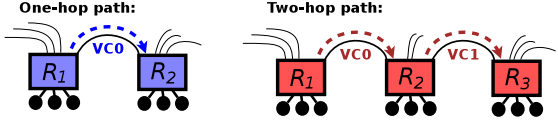}
\caption{Virtual channels in Slim Fly.}
\label{fig:vc}
\vspace{-1.5em}
\end{figure}

We use a strategy similar to the one introduced by Gopal~\cite{Gopal:1994:PSD:201173.201226,Duato:2002:INE:572400}.
We use two virtual channels (VC0 and VC1) for minimal routing. Assume we send a packet
from router $R_a$ to $R_b$. If the routers are directly connected, then
the packet is routed using VC0. If the path consists of two hops, then
the we use VC0 and VC1 for the first and the second hop, respectively.
We illustrate an example application of our strategy in Figure~\ref{fig:vc}.
Since the maximum distance in the network is two, only one turn can be
taken on the path and the number of needed VCs is thus no more than two.


For adaptive routing, we use four VCs (because
of the maximum number of turns with distance four). Here, we simply generalize
the scheme above and, for an $n$-hop path between $R_a$ to $R_b$,
we use a VC $k$ ($0 \le k < n$) on a hop $k$.

To avoid deadlocks in minimum routing one can also
use a generic deadlock-avoidance technique based on
automatic VC assignment to break cycles in the channel
dependency graph~\cite{Flich:2012:SET:2360762.2361136}. We tested the DFSSSP scheme
implemented in the Open Fabrics Enterprise Edition
(OFED)~\cite{domke-hoefler-dfsssp} which is available for generic
InfiniBand networks. OFED DFSSSP consistently needed three VCs to route
all \verb$SF$ networks. We also compared this number to random
\verb!DLN! networks~\cite{DBLP:conf/isca/KoibuchiMAHC12}, which
needed between 8 and 15 VLs for network sizes of 338 endpoints and 1,{}682
endpoints, respectively.

%
%
%

\section{Performance}
\label{performance}

\htor{What arbitration strategy?}

\goal{Say what we simulate}

In this section we evaluate the performance of MIN, VAL, UGAL--L, and
UGAL--G routing algorithms. We take into consideration various traffic
scenarios that represent the most important HPC workloads. 
First, we test uniform random traffic for graph computations, sparse linear algebra
solvers, and adaptive mesh refinement methods~\cite{Yuan:2013:NRS:2503210.2503229}.
Second, we analyze shift and permutation traffic patterns (bit complement, bit reversal, shuffle) that 
represent some stencil workloads and collectives such as all-to-all or all-gather~\cite{Yuan:2013:NRS:2503210.2503229}.
Finally, we evaluate a worst--case pattern designed specially for \verb$SF$ to test adversarial workloads.


\goal{Describe the simulation infrastructure and parameters}

%

We conduct
cycle-based simulations using packets that are injected with a Bernoulli process
and input-queued routers.
Before any measurements are taken, the simulator
is warmed up under load in order to reach steady-state. We use the
strategy in \cite{dally08} and utilize single flow control unit (flit)
packets to prevent the influence of flow control issues
(wormhole routing, virtual cut-through flow control) on the routing schemes.
Three virtual channels are used for each simulation.
Total buffering/port is 64 flit entries; we also simulated other
buffer sizes (8, 16, 32, 128, 256).
Router delay for credit processing is 2 cycles. Delays for channel latency, switch allocation,
VC allocation, and processing in a crossbar are 1 cycle each. 
Speedup of the internals of the routers over the channel transmission rate is 2. Input/output
speedups are set to 1.

%
%

\goal{Describe the topologies we compare to in simulations}

We compare topologies with full global bandwidth in Figure~\ref{fig:perf} and Sections~\ref{uniformTraffic}, \ref{shuffle}, \ref{WCScenario}, \ref{sec:buf_sizes}. We also provide results for
\emph{oversubscribed} \verb$SF$ in Section~\ref{sec:oversub}.
Due to space constraints and for clarity of plots we compare \verb$SF$ to two established topologies: Dragonfly (representing low-latency state-of-the-art networks) and fat tree (representing topologies offering high bisection-bandwidth). We select established and highly-optimized routing protocols for \verb$DF$ and \verb$FT-3$: \mbox{UGAL-L}~\cite{dally08} and the Adaptive Nearest Common Ancestor protocol (ANCA)~\cite{4228210}, respectively. We use \verb$FT-3$ instead of Long Hop since there is no proposed routing scheme for \verb$LH-HC$~\cite{2013arXiv1301.4177T} and designing such a protocol is outside the scope of our paper.
We present the results for $N$ $\approx$ 10K. Simulations of networks with $N \approx$ 1K, 2K, and 5K give similar results (latency varies by at most 10\% compared to networks with 10K nodes).
The parameters for \verb$DF$ are as follows: $k=27$, $p=7$, $N_r = 1,386$, $N = 9,702$. \verb$FT-3$ has $k=44$, $p=22$, $N_r = 1,452$, $N = 10,648$. Finally, \verb$SF$ has $k=44$, $p=15$, $N_r = 722$, $N = 10,830$. 
To enable fair performance comparison we simulate balanced variants of networks with full global bandwidth. Thus, they do not have exactly the same $N$;
we chose networks that vary by at most 10\% in $N$. We also investigated variants with exactly 10K endpoints that are either under- or oversubscribed; the
results follow similar performance patterns.
\verb$SF$ outperforms other topologies in terms of latency and offers comparable bandwidth.

\begin{figure*}
\centering
 \subfloat[Various buffer sizes.]{
  \includegraphics[width=0.189\textwidth]{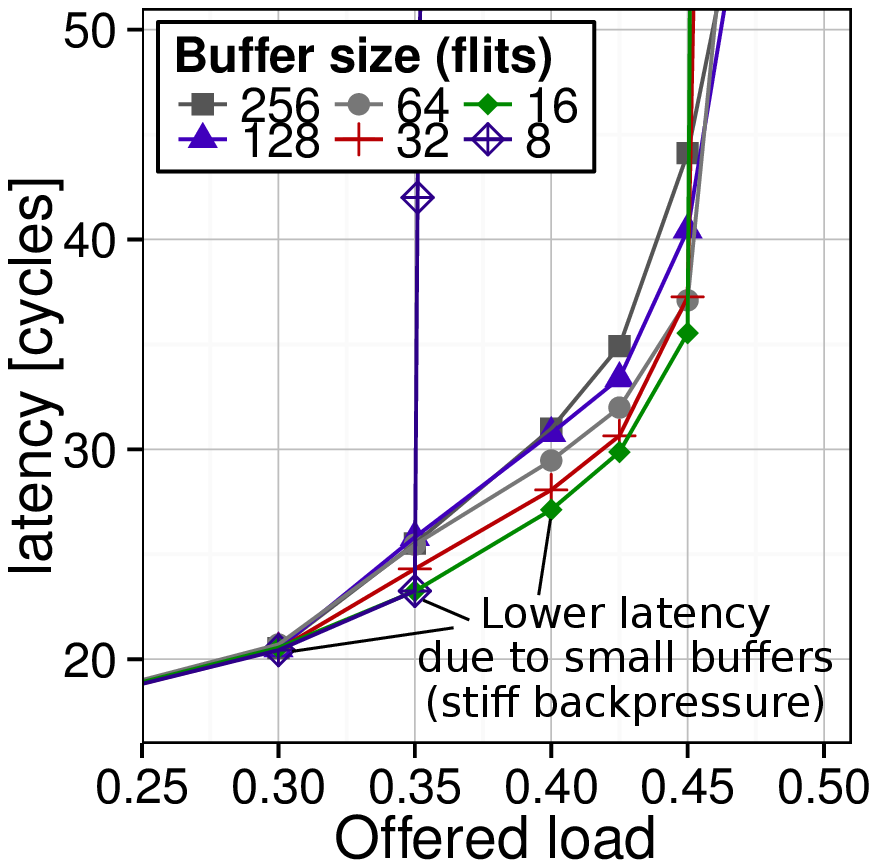}
  \label{fig:simSlimBadMMSBuffs}
 }
 \subfloat[Random traffic, $p=16$.]{
  \includegraphics[width=0.189\textwidth]{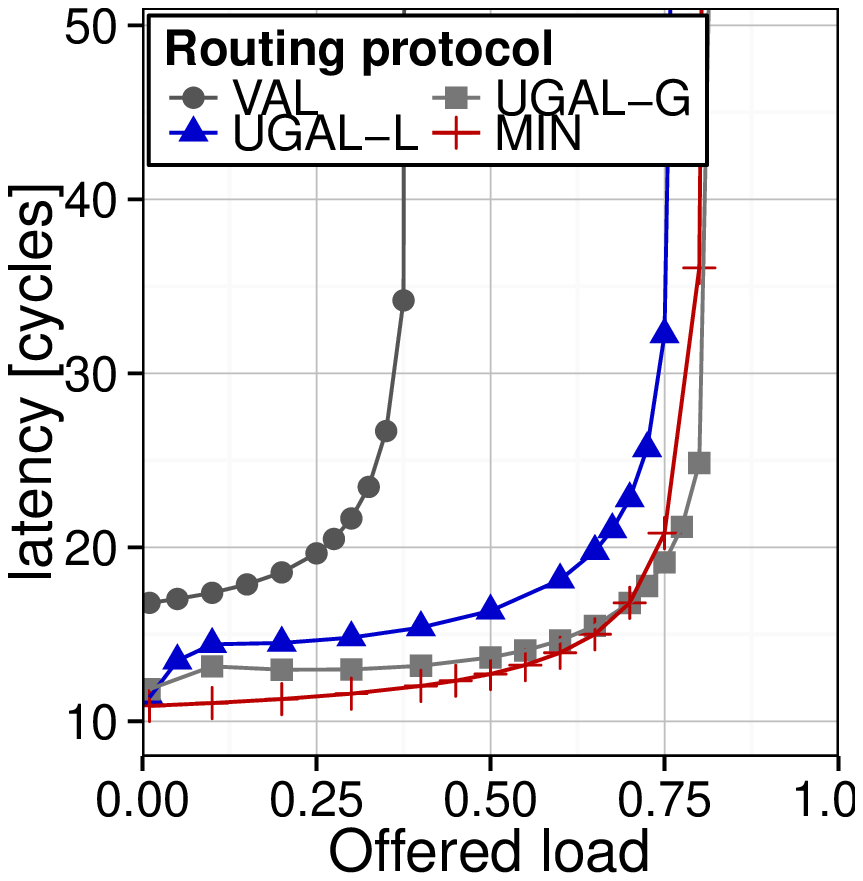}
  \label{fig:simSlimover-16_uniform}
 }	
 \subfloat[Worst-case traffic, $p=16$.]{
  \includegraphics[width=0.189\textwidth]{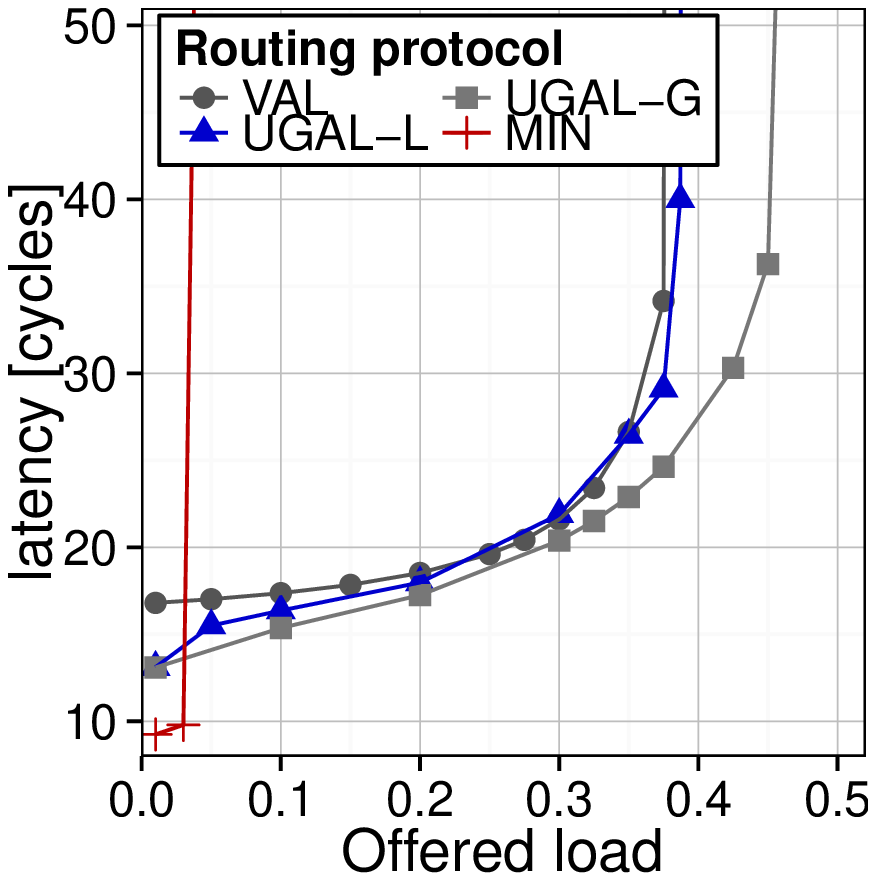}
  \label{fig:simSlimover-16_wc}
 }
 \subfloat[Random traffic, $p=18$.]{
  \includegraphics[width=0.189\textwidth]{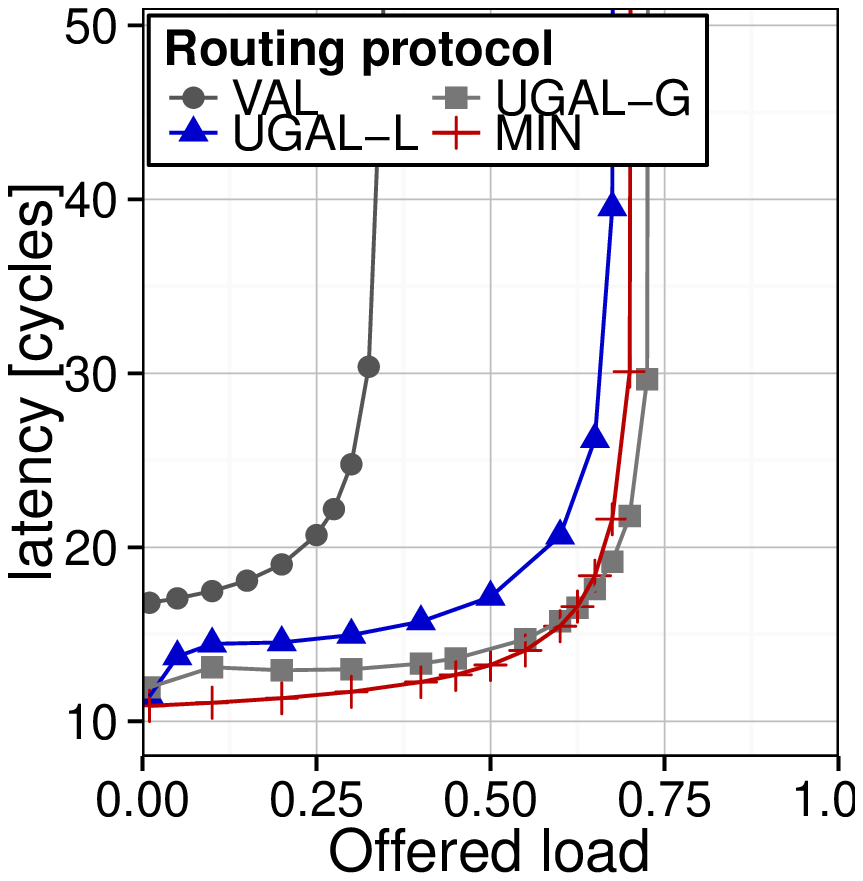}
  \label{fig:simSlimover-18_uniform}
 }
 \subfloat[Worst-case traffic, $p=18$.]{
  \includegraphics[width=0.189\textwidth]{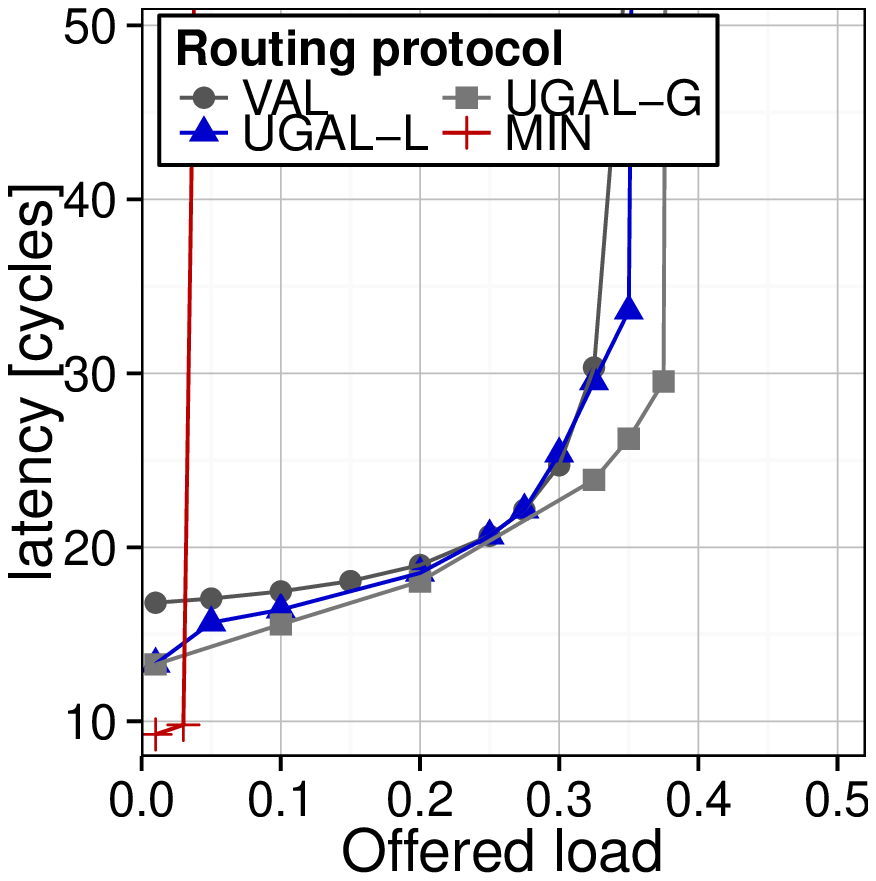}
  \label{fig:simSlimover-18_wc}
 }
\caption{Performance analysis of \texttt{SF}. In Figure~\ref{fig:simSlimBadMMSBuffs} we illustrate the influence of the router input buffer size on the performance of Slim Fly for the worst-case traffic. Figures~\ref{fig:simSlimover-16_uniform}~-~\ref{fig:simSlimover-18_wc} present the results of the simulations of different oversubscribed variants of \topologyName.}
\label{fig:perf_rest}
\end{figure*}

\subsection{Random Traffic for Irregular Workloads}
\label{uniformTraffic}

\goal{+Discuss random traffic}

In a random scenario each endpoint randomly selects the destination for
an injected packet. The results are presented in Figure~\ref{fig:simSlimUniformTraffic64}.
As expected, UGAL-G and MIN
achieve the best performance. VAL takes longer paths on average and
saturates at less than 50\% of the injection rate because it doubles the
pressure on all links. UGAL-L performs reasonably well (saturation at 80\% of the injection rate) but packets take
some detours due to transient local backpressure. This slightly decreases the
overall performance at medium load but converges towards full bandwidth
for high load (the difference is around 5\% for the highest injection
rate; this effect, described in~\cite{Jiang:2009:IAR:1555754.1555783}, is
much less visible in \verb$SF$ than in \verb$DF$ thanks to \verb$SF$'s lower diameter resulting in fewer queues that can congest).
As expected from Figure~\ref{fig:BB}, \verb$DF$ offers lower bandwidth
while the bandwidth of \verb$FT-3$ is slightly higher than \verb$SF$.
Finally, \verb$SF$ has the lowest latency due to its lower $D$ than in \verb$DF$
and \verb$FT-3$.


\subsection{Bit Permutation and Shift Traffic for Collective Operations}
\label{shuffle}

\goal{+Discuss bit permutation scenarios that we simulate}

We use several bit permutation scenarios to fully evaluate the
performance of \verb$SF$.
As $N$ has to
be a power of two we artificially prevent some endpoints from sending
and receiving packets for the purpose of this evaluation. The number of
endpoints that are active is 8,{}192 (power of two closest to the original size of the networks).
We denote $b$ as the number of bits in the endpoint address, $s_i$ as the $i$th bit of the source endpoint address, and $d_j$ as the $j$th bit of the destination endpoint address. 
We simulate the shuffle ($d_i = s_{i-1~mod~b}$), bit reversal ($d_i = s_{b-i-1}$), and bit complement ($d_i = \neg s_i$) traffic pattern. We also evaluate a shift pattern in which, for source endpoint $s$, destination $d$ is (with identical probabilities of $\frac{1}{2}$) equal to either $d = (s\;mod\;\frac{N}{2}) + \frac{N}{2}$ or $d = (s\;mod\;\frac{N}{2})$. 
%
%
We present the results in Figures~\ref{fig:simSlimBitrevTraffic64}--\ref{fig:simSlimAsymTraffic64} (due to space constraints we skip bit reverse/complement). The bandwidth of \verb$FT-3$, higher than UGAL--L and only slightly better than UGAL--G, indicates that the local decisions made by UGAL-L miss some opportunity for traffic balancing. As expected, \verb$SF$ offers slightly higher bandwidth and has lower latency than \verb$DF$.

\subsection{Worst--Case Traffic for Adversarial Workloads}
\label{WCScenario}

\goal{+Explain the WC traffic patterns we use}

\begin{figure}[h!]
\centering
\includegraphics[width=0.4\textwidth]{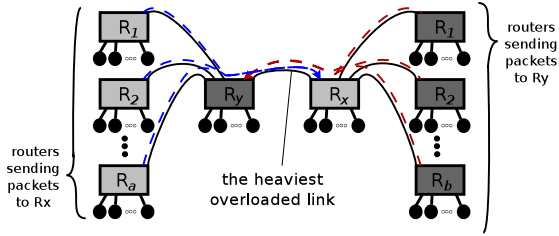}
\caption{Illustration of the worst--case scenario for \topologyName.}
\label{fig:wcPattern}
\end{figure}

We now describe the worst-case traffic pattern for minimal deterministic
routing on Slim Fly networks. For this, we consider only traffic patterns
that do not overload endpoints. The scheme is
shown in Figure~\ref{fig:wcPattern}. The worst-case pattern for a
Slim Fly network is when all $p$ endpoints attached to routers $R_1, ..., R_a$
send and receive from all endpoints at router $R_x$ and the shortest
path is of length two and leads via router $R_y$. In addition, all $p$
endpoints at routers $R_1, ..., R_b$ send and receive from all endpoints at
router $R_y$ and the shortest path leads through router $R_x$. This puts
a maximum load on the link between routers $R_x$ and $R_y$.
We generate this pattern by selecting a link between $R_x$ and $R_y$ and choosing routers
$R_1, ..., R_a$ and $R_1, ..., R_b$ according to the description above until all
possibilities are exhausted. For \verb$DF$ we use a worst-case traffic
described in Section~4.2 in~\cite{dally08}. In \verb$FT-3$ we utilize a pattern
where every packet traverses core (highest-level) switches in the topology.

\begin{figure*}
\centering
\includegraphics[width=1.0\textwidth]{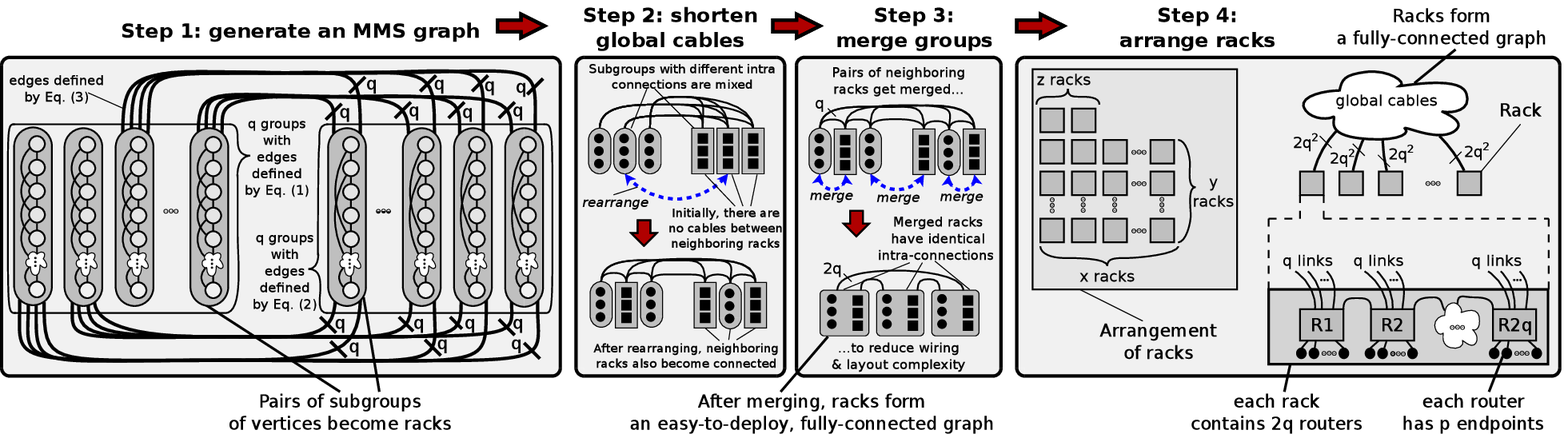}
\caption{An MMS graph and the corresponding datacenter layout.}
\label{fig:mms_to_layout}
\end{figure*}

\goal{+Discuss the results}

Figure~\ref{fig:simSlimWCTraffic64} shows the simulation results of
adversarial traffic. MIN routing is limited to $\frac{1}{p+1}$ throughput in the worst-case.
VAL and UGAL-L can disperse the traffic across
multiple channels and can support up to 40\% (VAL) and 45\% (UGAL-L)
offered load, providing slightly higher bandwidth than \verb$DF$. As we use the balanced full-bandwidth
variant of \verb$FT-3$, it achieves higher bandwidth than both \verb$DF$ and \verb$SF$.

\subsection{Study of Buffer Sizes}
\label{sec:buf_sizes}

\goal{+Discuss the study on buffer sizes}

We also analyze how the size of input router buffers affects the performance of \verb$SF$. We present the results for the worst-case traffic in Figure~\ref{fig:simSlimBadMMSBuffs} (other scenarios follow similar performance patterns). Smaller sizes result in lower latency (due to stiffer backpressure propagation), while bigger buffers enable higher bandwidth.

\subsection{Oversubscribing Slim Fly Networks}
\label{sec:oversub}

\goal{+Explain oversubscription and its goals}

Oversubscribing
the number of endpoints per router increases the flexibility
of port count and cost of \verb$SF$. We define an
oversubscribed network as a network which cannot achieve full global
bandwidth, cf. Section~\ref{attachingEndpoints}.

\goal{+Discuss the results}

Figures~\ref{fig:simSlimover-16_uniform}--\ref{fig:simSlimover-18_wc} show the latency and bandwidth of
different oversubscribed \verb$SF$ networks with network radix $k'=29$. In its full-bandwidth configuration
($p=15$) it supports 10,{}830 endpoints. We investigate six different
oversubscribed networks with concentration 16--21
connecting from 11,{}552 up to 15,{}162 endpoints, respectively. We present the results for $p=16$ and $p=18$, other cases follow similar performance patterns. According
to~\cite{Dally:2003:PPI:995703}, we define the accepted bandwidth as the
offered load of random uniform traffic which saturates the network.
The full-bandwidth \verb$SF$ can accept up to 87,5\% of the
traffic. The \verb$SF$ with $p=16$ and $p=18$ accept up to
80\% and 75\% of the offered traffic, respectively. The bandwidth for the worst-case
traffic behaves similarly. This study illustrates the flexibility of the \verb$SF$ 
design that allows for adding new endpoints while
preserving high bandwidth and low latency.



\goal{+State final evaluation conclusions}

We conclude that \verb$SF$ can deliver lower latency and in most cases comparable bandwidth in comparison to
other topologies. As we will show in Section~\ref{costComparison}, by lowering the diameter Slim Fly offers 
\emph{comparable bandwidth and lower latency for lower price and energy consumption per endpoint}.

\section{Cost and Power Comparison}
\label{costComparison}

\goal{Introduce the section}

We now proceed to provide cost and power comparison of \verb$SF$ with
other topologies. We also discuss the engineering constraints and
partitioning of \verb$SF$ into groups of routers.

\subsection{Physical Layout}
\label{arrangementInCabinets}

\goal{+State partitioning of networks is challenging and say what we focus on}

One engineering challenge for a low-diameter network is how to
arrange it in an HPC center or a datacenter with minimal cabling costs.
We now describe a possible physical arrangement of \verb$SF$. We focus on making \verb$SF$ deployable (with
symmetric partitioning/modularity). Remaining issues such as incorporating power supply units can be solved with
well-known strategies used for other modular networks (e.g., \verb$DF$).

\goal{+Discuss the general partitioning idea, based on MMS structure}


We arrange the routers and their attached endpoints into
racks with an equal number of cables connecting the racks. We partition Slim Fly basing on the modular structure of the underlying MMS graph (see Section~\ref{diam2Construct}, Figure~\ref{fig:mms_general}, and the left side of Figure~\ref{fig:mms_to_layout} (Step~1)). The MMS modular design enables several different ways of easy partitioning. We focus here on the most intuitive one, valid for prime $q$: two corresponding subgroups of vertices (one consisting of routers $(0,x,y)$, the other consisting of routers $(1,m,c)$) form one rack. The $q$ connections between these two subgroups plus their original intra-group edges defined by Equations~(1) and~(2) become intra-group cables of a single rack.

\goal{+Discuss the first stage of partitioning}

We illustrate how a datacenter layout originates from an MMS graph in Figure~\ref{fig:mms_to_layout}. First, in order to limit the cost, we rearrange subgroups of routers so that the length of global cables is reduced (Step~2). Note that, from the point of view of the MMS structure, we simply utilize the fact that no edges connect subgroups of routers $(0,x,y)$ with one another (the same holds for  routers $(1,m,c)$).

\goal{+Discuss the second stage of partitioning}

\sloppy
Second, the neighboring groups of routers $(0,x,y)$ and $(1,m,c)$ are merged; newly-created groups of vertices form racks (Figure~\ref{fig:mms_to_layout}, Step~3). Note that, as we always merge one group of routers $(0,x,y)$ with another group of routers $(1,m,c)$, after this step each rack has \emph{the same pattern of intra-group cables}. In addition, the whole datacenter can now be viewed as a fully-connected graph of identical racks, with $2q$ inter-connections between every pair of racks.
Such a design facilitates the wiring and datacenter deployment.

\goal{+Discuss the final stage of partitioning}

The final layout is illustrated in Step~4 in Fig.~\ref{fig:mms_to_layout}. We place the racks as a square (or a rectangle close to a square) where $x$ and $y$ are the numbers of racks along the corresponding
dimensions. If the number of racks $N_{rck}$ is not divisible by any $x$ and $y$, then we find $z$ such that $N_{rck} = x \cdot y + z$ and we place remaining $z$ racks at an arbitrary side. 

\begin{figure*}
\centering

 \subfloat[Cable cost model.]{
  \includegraphics[width=0.153\textwidth]{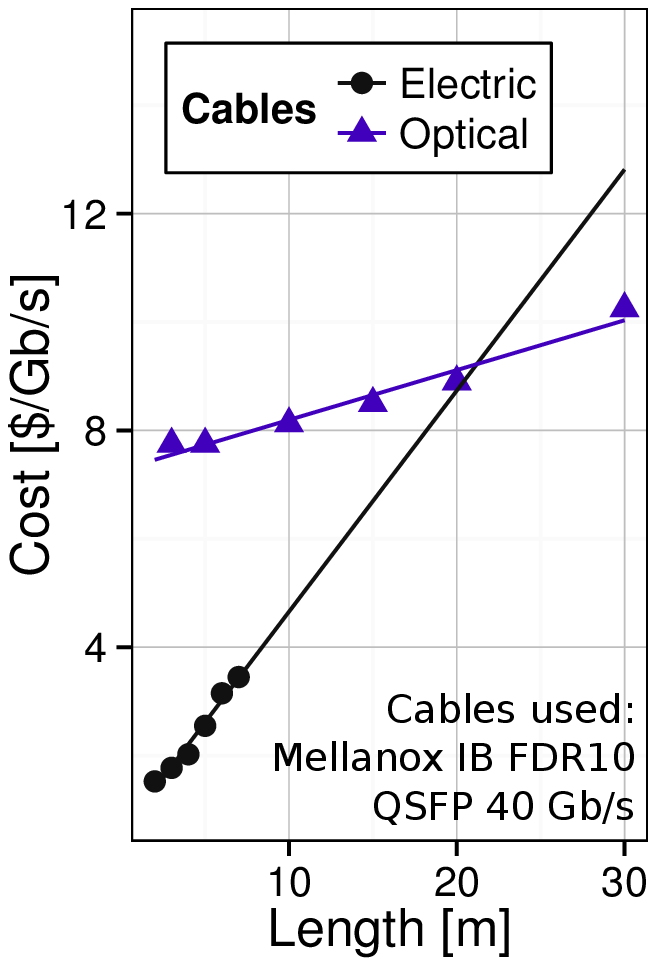}
  \label{fig:cablesCost}
 }\hfill
 \subfloat[Routers cost model]{
  \includegraphics[width=0.153\textwidth]{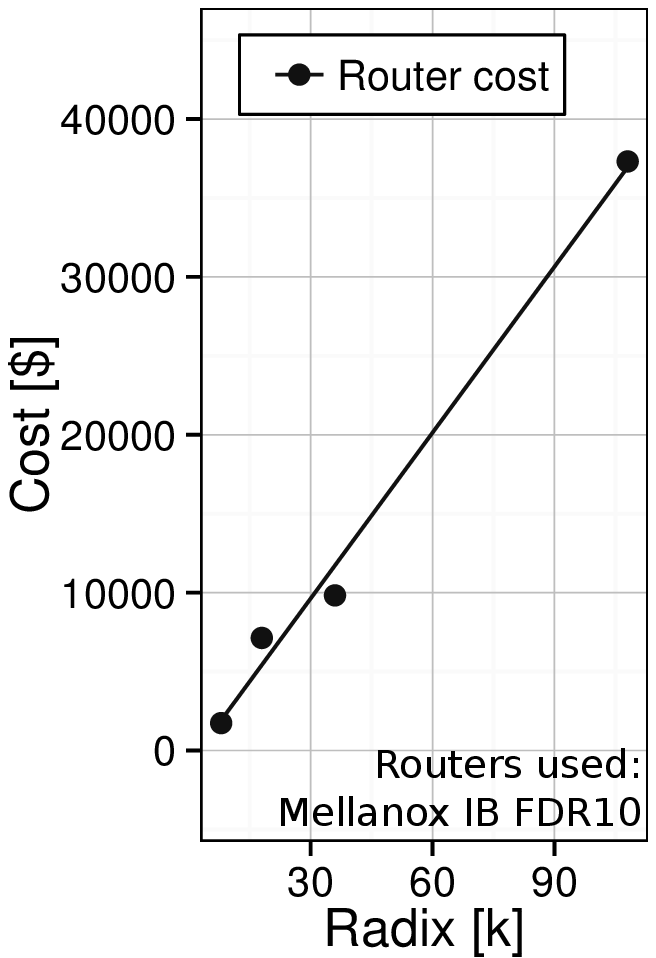}
  \label{fig:routersCost}
 }
 \hfill 
 \subfloat[Total cost of the network.]{
  \includegraphics[width=0.32\textwidth]{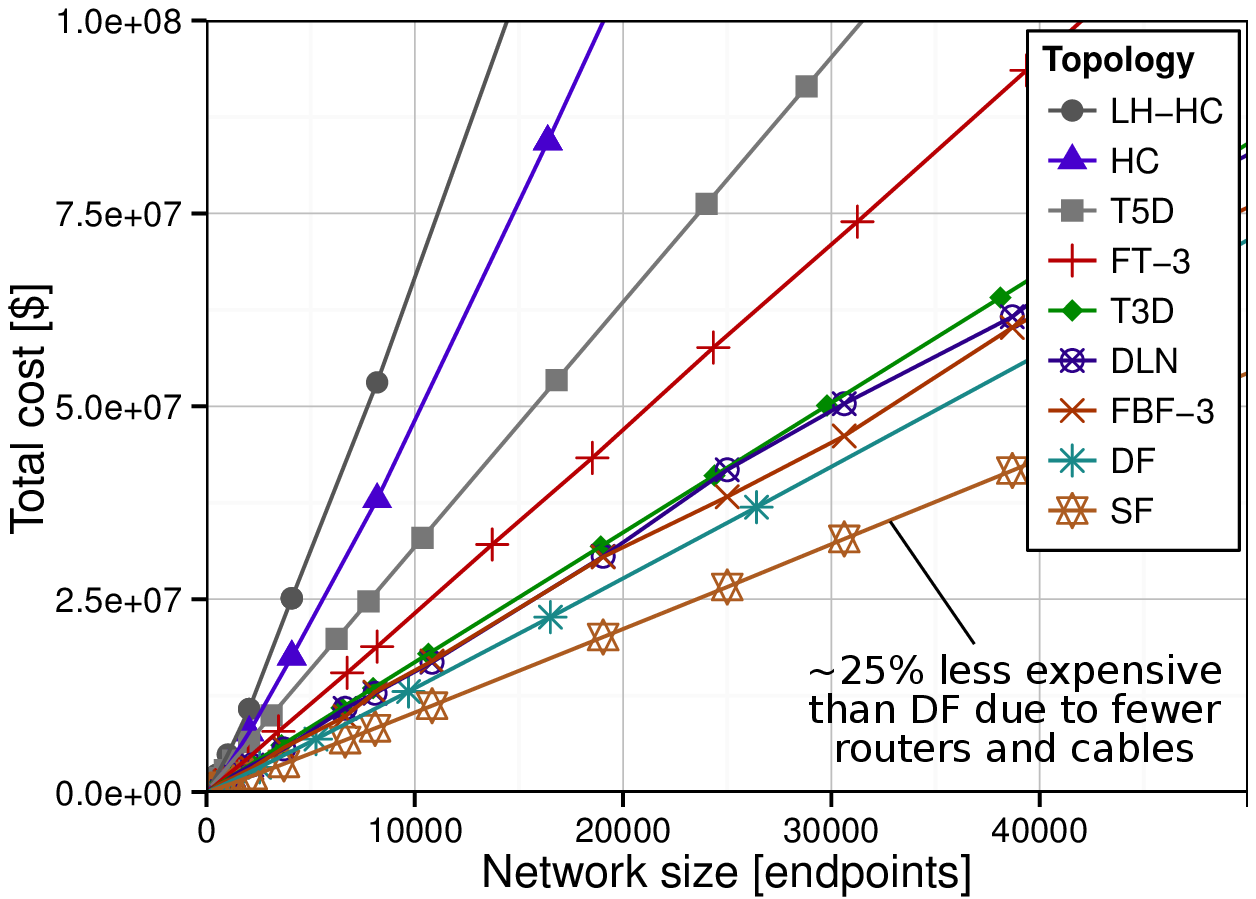}
  \label{fig:costPerNode}
 }
  \hfill 
 \subfloat[Total power consumed by the network.]{
  \includegraphics[width=0.32\textwidth]{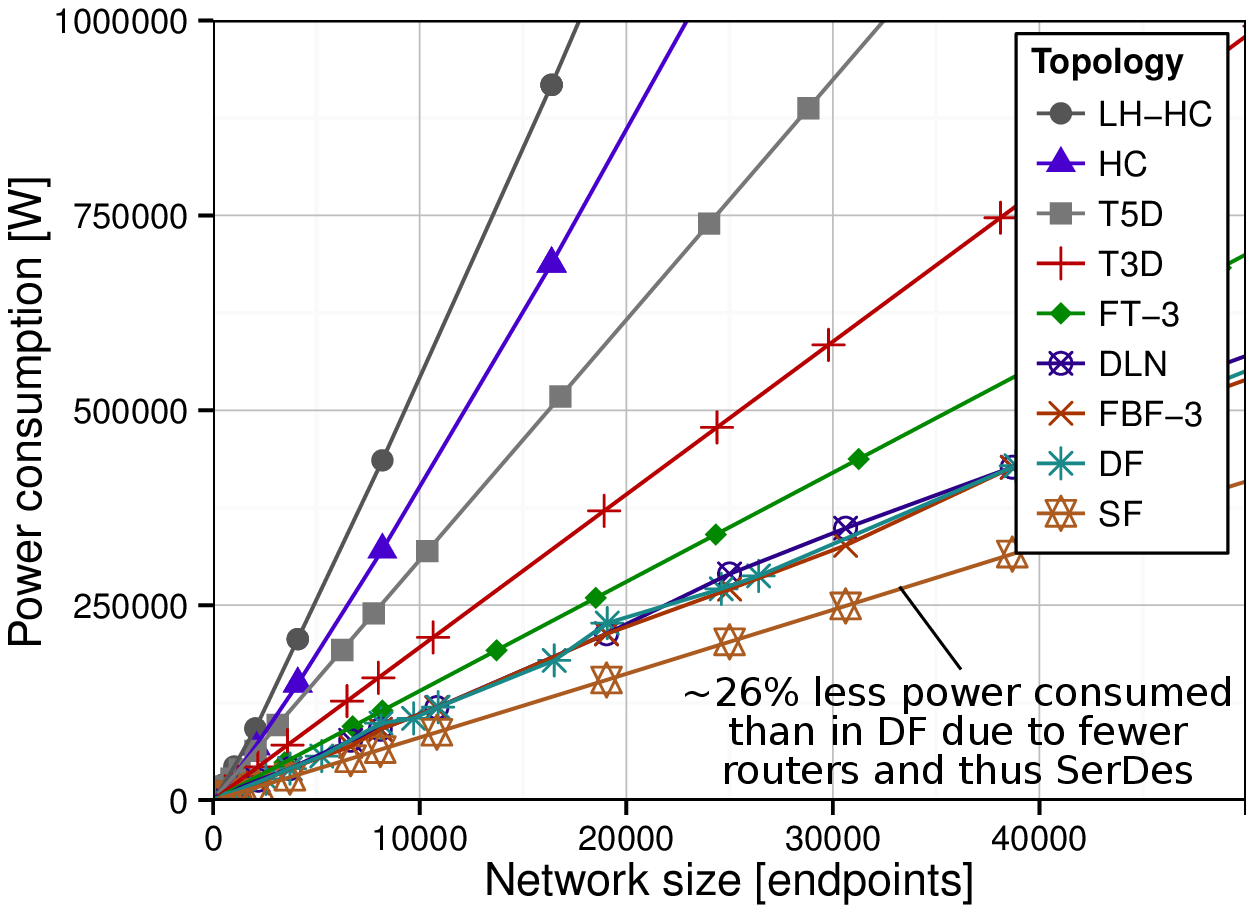}
  \label{fig:power}
 }
  \caption{The details of the cost \& power model and the comparison of Slim Fly to other topologies (cables: Mellanox InfiniBand (IB) FDR10 40Gb/s QSFP, routers: Mellanox IB FDR10).}
\label{fig:costModelDetails}
\end{figure*}

\begin{figure*}
\centering

 \subfloat[Cable cost model.]{
  \includegraphics[width=0.153\textwidth]{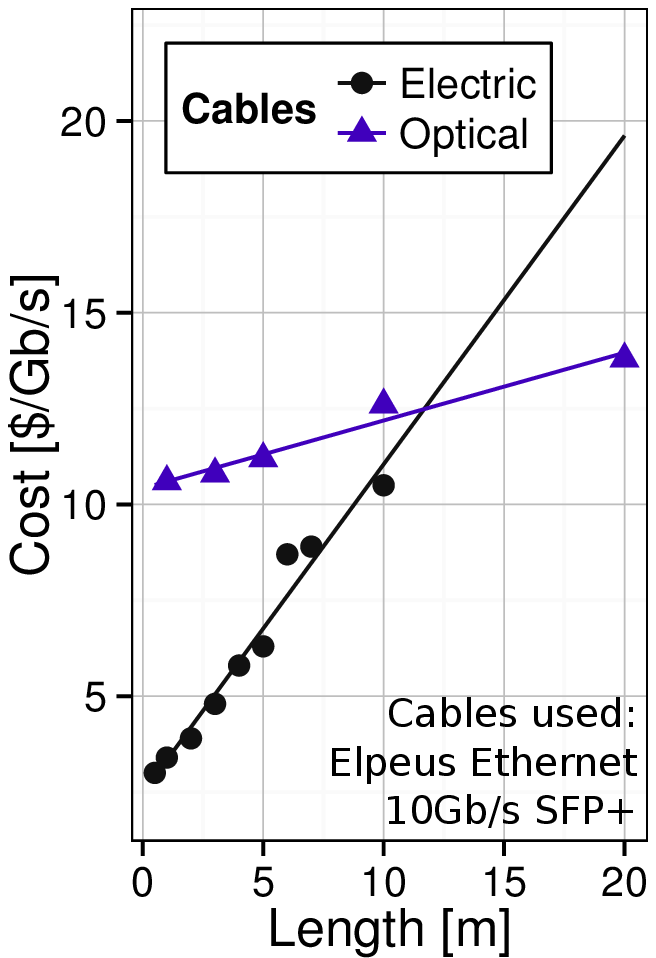}
  \label{fig:cablesCost}
 }\hfill
 \subfloat[Routers cost model]{
  \includegraphics[width=0.153\textwidth]{routers_labels.eps}
  \label{fig:routersCost}
 }
 \hfill 
 \subfloat[Total cost of the network.]{
  \includegraphics[width=0.32\textwidth]{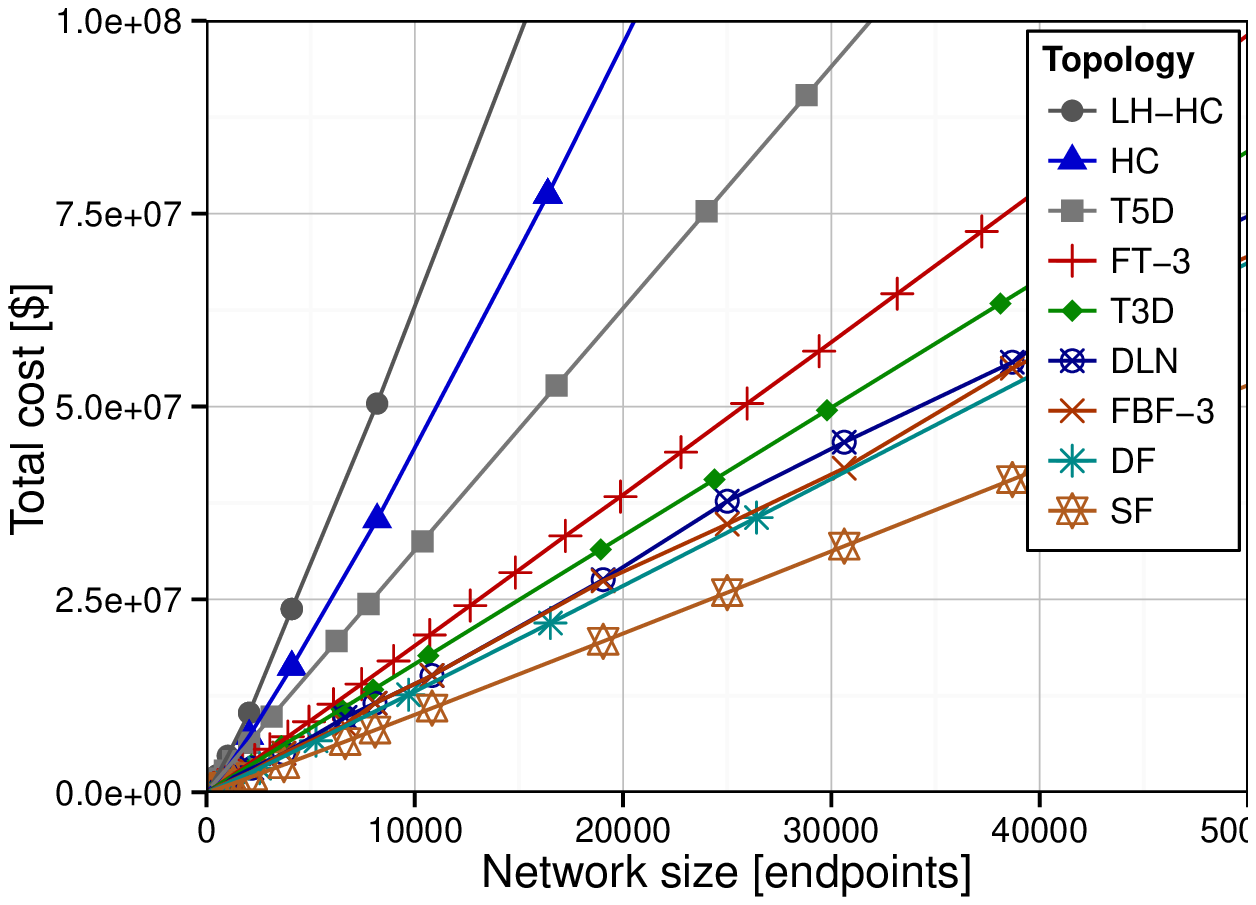}
  \label{fig:costPerNode}
 }
  \hfill 
 \subfloat[Total power consumed by the network.]{
  \includegraphics[width=0.32\textwidth]{power_labels.eps}
  \label{fig:power}
 }
  \caption{The details of the cost \& power model and the comparison of Slim Fly to other topologies (cables: Elpeus Ethernet
10Gb/s SFP+, routers: Mellanox IB FDR10).}
\label{fig:costModelDetailsElpeusCables}
\end{figure*}

\begin{figure*}
\centering

 \subfloat[Cable cost model.]{
  \includegraphics[width=0.153\textwidth]{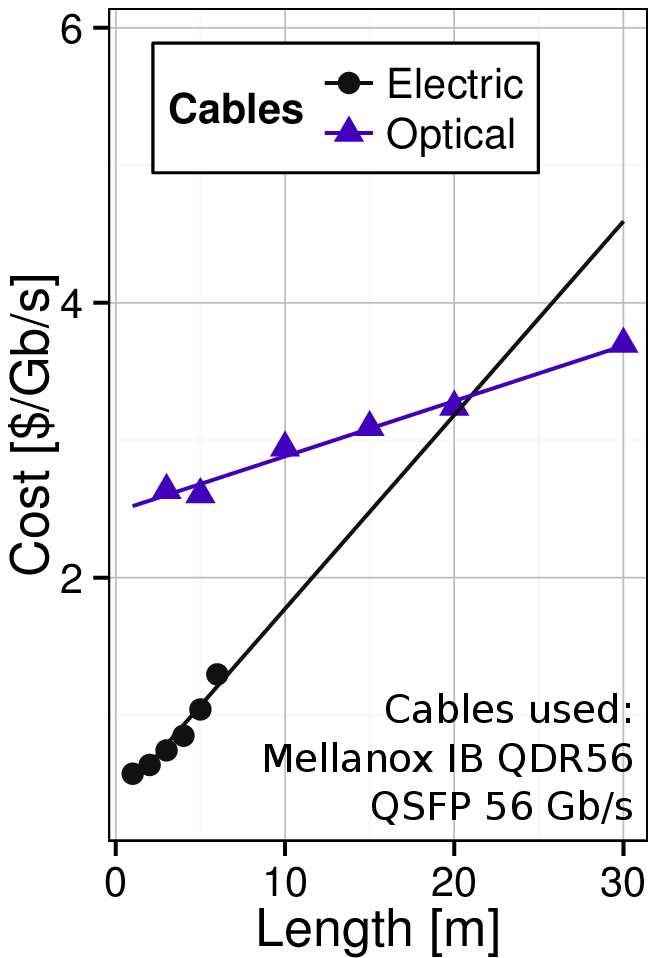}
  \label{fig:cablesCost}
 }\hfill
 \subfloat[Routers cost model]{
  \includegraphics[width=0.153\textwidth]{routers_labels.eps}
  \label{fig:routersCost}
 }
 \hfill 
 \subfloat[Total cost of the network.]{
  \includegraphics[width=0.32\textwidth]{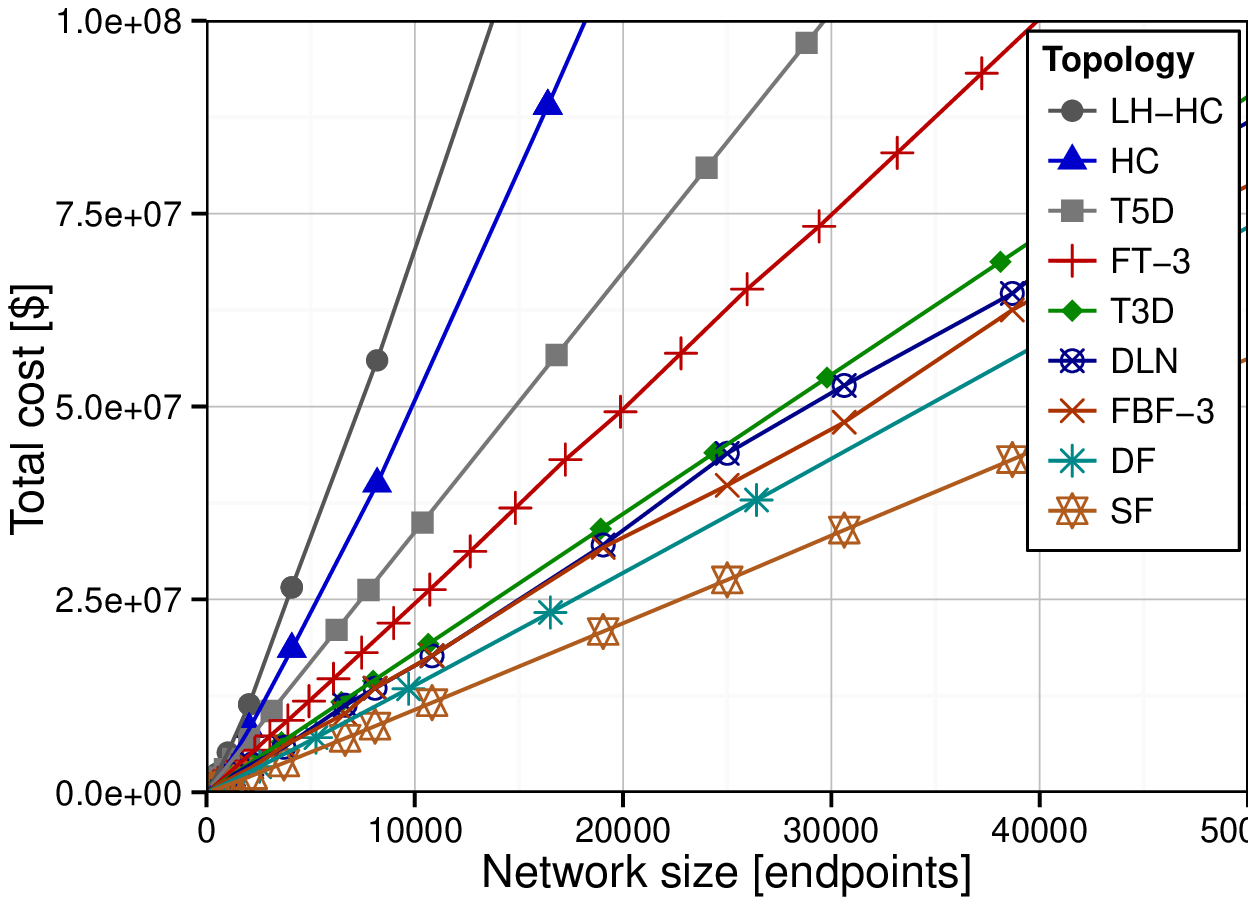}
  \label{fig:costPerNode}
 }
  \hfill 
 \subfloat[Total power consumed by the network.]{
  \includegraphics[width=0.32\textwidth]{power_labels.eps}
  \label{fig:power}
 }
  \caption{The details of the cost \& power model and the comparison of Slim Fly to other topologies (cables: Mellanox InfiniBand (IB) QDR56 56Gb/s QSFP, routers: Mellanox IB FDR10).}
\label{fig:costModelDetailsQDR56Cables}
\end{figure*}

\goal{+Discuss an example partitioning}

As an example, consider an SF MMS network with $q=19$, consisting of 10,{}830 endpoints, with router radix $k' = 29$, concentration $p \approx \left\lceil
k'/2\right\rceil = 15$ and $k = k' + p = 44$. For this network, we have $q = 19$ racks, each containing 38 routers (570 endpoints), and 38 global channels to every other group. A different layout would allow for $q=39$ racks with 19 routers and 285 endpoints in each rack.





\subsubsection{Slim Fly Layout vs. Dragonfly Layout}
\label{SFvsDFLayout}

\goal{++Describe differences in layout of SF and DF}

The final layout of \verb$SF$ is similar to that of \verb$DF$: both form a 2-level hierarchy consisting of routers and groups of routers. We propose such construction scheme to facilitate the reasoning about \verb$SF$. There are still some differences between \verb$SF$ and \verb$DF$ that ensure lower diameter/higher resiliency in \verb$SF$:

\begin{itemize}[leftmargin=1em]
\item Routers inside each group in \verb$DF$ constitute fully-connected graphs. Routers inside groups in \verb$SF$ are not necessarily fully-connected.
\item In \verb$DF$, every router is connected to all $a-1$ remaining local routers in a group; in \verb$SF$ every router is connected to $\frac{a-\delta}{2}+1$ other local routers, which means that there are $\approx$50\% fewer cables in a \verb$SF$ router group than in a \verb$DF$ router group.
\item In \verb$DF$, there is one inter-group cable connecting two groups. In \verb$SF$, two groups are connected using $2q$ cables.
\item A balanced \verb$SF$ has higher concentration ($p \approx 33\%k$) than a balanced same-size \verb$DF$ ($p \approx 25\%k$).
This results in higher endpoint density and $\approx$25\% fewer routers/racks in \verb$SF$.
\end{itemize}

\subsection{Cost Model}


\goal{+Introduce the section and describe the model}

We now describe a cost model (similar to the model
used in~\cite{dally07}) that includes the cost of routers and interconnection cables,
which usually constitute the vast majority of the overall network
costs~\cite{dally07}. We assume
that routers together with endpoints are grouped in racks of size $1
\times 1 \times 2$ meters. Local (intra-rack) links are electric while
global (inter-rack) channels are optic.
Routers are placed on top of racks.
The maximum Manhattan distance between two routers in a rack is
$\approx$2m and the minimum is 5-10cm, thus on average
intra-rack cables are 1m long.
The distance between two racks is also calculated using the Manhattan
metrics. Following~\cite{dally07}, we add 2 meters of cable
overhead for each global link. Racks are arranged in a shape close to a square
as presented in Section~\ref{arrangementInCabinets}.

\subsubsection{Cables}

\goal{++Describe the details of cables/routers pricing}

To estimate the cost of network cables we use data
bandwidth as a function of distance (in meters).
We apply linear regression to today's pricing data\footnote{Prices are based on http://www.colfaxdirect.com} to get the
cost functions. We use Mellanox InfiniBand (IB) FDR10 40Gb/s QSFP cables. Cost of electrical
cables can be estimated as $f(x) = 0.4079x + 0.5771$ [\$/Gb/s], while for optical fiber channels
we have $f(x)=0.0919x + 2.7452$ [\$/Gb/s]. Figure~\ref{fig:cablesCost} shows the model. Other cables that we considered are Mellanox IB QDR 56Gb/s QSFP,
Mellanox Ethernet 40Gb/s QSFP, Mellanox Ethernet 10Gb/s SFP+, and Elpeus Ethernet 10Gb/s SFP+.
They result in similar cost patterns (final relative cost differences between topologies vary by $\approx$1-2\%; see Figure~\ref{fig:costModelDetailsElpeusCables} and~\ref{fig:costModelDetailsQDR56Cables}).

\subsubsection{Routers}

We also provide a function to calculate router cost basing on state of the
art Mellanox IB FDR10 routers. 
We assume router cost to
be a linear function of the radix because the router chip often has a
rather constant price which is mainly determined by the development
costs~\cite{dally08} while the SerDes are often the most
expensive part of a router. We use linear regression to calculate the fit
($f(k) = 350.4k - 892.3$~[\$]) and we show the model in Figure~\ref{fig:routersCost}.
Other tested routers are Mellanox Ethernet 10/40Gb, they again only negligibly impact the 
relative cost differences between topologies ($\approx$1\% difference between \verb$SF$ and \verb$DF$).

%
%

\subsubsection{Models of Remaining Network Topologies}

\paragraph{Tori}
\label{sec:torus_cost_model}

We model \verb$T3D$ and \verb$T5D$ as cuboids and hyper cuboids, respectively.
Following~\cite{dally08} we assume that tori have folded design that do not require optical links.

\paragraph{Hypercube and Long Hop}
\label{sec:hypercube_cost_model}

In \verb$HC$ and \verb$LH-HC$, we use electric cables for
intra- and fiber cables for inter-rack connections.
Each router connects to a single router in each
dimension.
In \verb$LH-HC$ routers have additional $L$ ports
to other routers as specified in Section E-S-3 of~\cite{2013arXiv1301.4177T}.

\paragraph{Fat tree}

\verb$FT-3$ has 3 layers with the sum of $5 p^2$ routers that are
installed in a central row in the network. Core routers
are connected to aggregation routers with $2 p^3$ optical cables. 
Each aggregation router is connected to $p$ edge
routers giving a total of further $2 p^3$ fiber channels. We estimate
average cable length between routers to be 1m.
Finally, the number of endpoints and the cables connecting them
to routers is also $2 p^3$;  we assume that links shorter than 20 meters are electrical.
$p^2$ endpoints form a single group (pod).

\paragraph{Flattened butterfly}

We arrange routers and groups in \verb$FBF-3$ as
in~\cite{dally07}. There are $p$ routers in every group
(rack) and $p^2$ groups forming an ideal square. Each
group is fully connected ($\frac{p (p - 1)}{2}$ electric channels) and
there are $p$ fiber links between every two groups in the same row or
column of racks.

\paragraph{Dragonfly and Random Networks}

We use the balanced \verb$DF$ \cite{dally08} ($a = 2p =
2h$). $a$ is
the number of routers in a group and $h$ is the number of fiber
cables connected to each router. There are $g = a \cdot h + 1$ fully
connected groups of routers, each having $\frac{a  (a - 1)}{2}$ electric
cables. Groups form a clique with the total of  $\frac{g
(g - 1)}{2}$ fiber cables~\cite{dally08}. \verb$DLN$ have groups
with the same size ($a$), but cables are placed randomly.

\begin{table*}[ht]
\centering
\footnotesize
\setlength{\tabcolsep}{5.0pt}
\begin{tabular}{@{}l|cccc|cccc|cccc|c|c@{}}
\toprule
   & \multicolumn{4}{c}{Low-radix topologies}           & \multicolumn{10}{|c}{High-radix topologies}                                                                                                     \\ \midrule
Topology           & \verb$T3D$ & \verb$T5D$ & \verb$HC$ & \verb$LH-HC$ & \verb$FT-3$ & \verb$DLN$ & \verb$FBF-3$ & \verb$DF$ & \verb$FT-3$ & \verb$DLN$ & \verb$FBF-3$ & \verb$DF$ & \verb$DF$        & \textbf{\ttfamily{SF}}        \\ \midrule
Endpoints ($N$)    & \textbf{10,648}     & \textbf{10,368}     & \textbf{8,192}     & \textbf{8,192}        & 19,876      & 40,200     & 20,736       & 58,806    & \textbf{10,718}      & \textbf{9,702}      & \textbf{10,000}       & \textbf{9,702}     & \textbf{10,890}           & \textbf{10,830}            \\
Routers ($N_r$)    & 10,648     & 10,368     & 8,192     & 8,192        & 2,311       & 4,020      & 1,728        & 5,346     & 1,531       & 1,386      & 1,000        & 1,386     & 990              & \textbf{722}              \\
Radix ($k$)        & 7          & 11         & 14        & 19           & \textbf{43}          & \textbf{43}         & \textbf{43}           & \textbf{43}        & 35          & 28         & 33           & 27        & \textbf{43}               & \textbf{43}               \\
Electric cables    & 31,900     & 50,688     & 32,768    & 53,248       & 19,414      & 32,488     & 9,504        & 56,133    & 7,350       & 6,837      & 4,500        & 9,009     & 6,885            & \textbf{6,669}             \\
Fiber cables       & 0          & 0          & 12,288    & 12,288       & 40,215      & 33,842     & 20,736       & 29,524    & 24,806      & 7,716      & 10,000       & 4,900     & 1,012            & \textbf{6,869}             \\ \midrule
Cost per node [\$] & 1,682      & 3,176      & 4,631     & 6,481        & 2,346       & 1,743      & 1,570        & 1,438     & 2,315       & 1,566      & 1,535        & 1,342     & 1,365            & \textbf{1,033}             \\
Power per node [W] & 19.6       & 30.8       & 39.2      & 53.2         & 14.0        & 12.04      & 10.8         & 10.9      & 14.0        & 11.2       & 10.8         & 10.8      & 10.9             & \textbf{8.02}              \\ \bottomrule
\end{tabular}
\caption{Cost and power comparison between a Slim Fly ($N=10830, k=43$) and other networks (\cref{sec:cost_results} and \cref{sec:power_results}). We select low-radix networks with $N$ comparable to that of Slim Fly. $N$ cannot be identical due to the limited number of existing network configurations. For high-radix topologies, we select comparable $N$ and we also compare to topologies with fixed radix $k$. We also construct and analyze one additional variant of a DF that has both comparable $N$ \emph{and} identical $k$ as the analyzed SF. Each of these groups of topologies is indicated with a bolded parameter.}
\label{tab:detailed_cost_study}
\end{table*}

\subsubsection{Discussion of the Results}
\label{sec:cost_results}

\goal{++Present the cost results}

Figure~\ref{fig:costPerNode} presents the total cost of balanced networks. A detailed case-study showing cost per endpoint for an \verb$SF$ with $\approx$10K endpoints and radix $43$ can be found in Table~\ref{tab:detailed_cost_study}. Here, we first compare \verb$SF$ to
low-radix topologies (\verb$T3D$, \verb$T5D$, \verb$HC$, \verb$LH-HC$) with comparable network size $N$. $N$ cannot be identical for each topology due to the limited number of networks in their balanced configurations. We use tori with size close to that of \verb$SF$ (1-4\% of difference). However, a small number of \verb$HC$ and \verb$LH-HC$ configurations forced us to use $N=8,192$ for these topologies. We additionally constructed \emph{hybrid} hypercubes and Long Hops that consist of excessive routers and endpoints and are thus identical in size to \verb$SF$; the cost results vary by only $\approx$1\%. \verb$LH-HC$ is more expensive than \verb$HC$ because it uses additional links to increase bisection bandwidth. \verb$SF$ is significantly more cost-effective than low-radix networks as it uses fewer routers and cables.

Next, we present the results for balanced high-radix networks (\verb$FT-3$, \verb$DLN$, \verb$FBF-3$, \verb$DF$). We first compare to topologies that have similar $N$ (at most 10\% of difference). Then, we select networks with the same radix $k$ as the analyzed \verb$SF$. We also compare to one additional variant of a \verb$DF$ that has both comparable $N$ \emph{and} identical $k$ as the analyzed \verb$SF$. Such a construction is possible for \verb$DF$ because it has flexible structure based on three parameters $a$, $h$, and $p$ that can have any values. We perform an exhaustive search over the space of all Dragonflies that satisfy the condition $a \ge 2h $ and $p \ge h$. This condition ensures full utilization of global channels (see Section~3.1 in~\cite{dally08} for details). We select a \verb$DF$ that has $k=43$ and whose $N$ is closest to that of the analyzed \verb$SF$. 
In all cases, \verb$SF$ is $\approx$25\% more
cost-effective than \verb$DF$, and almost 30\%, 40\%, and 50\% less expensive than \verb$FBF-3$, \verb$DLN$, and \verb$FT-3$.
The difference between \verb$SF$ and other topologies is achieved by the reduction in the number of
needed routers and cables and the today's commodization of fiber optics. For example, for a network with $k=43$ and $N \approx 10,000$, \verb$DF$ uses 990 routers while \verb$SF$ utilizes only 722 routers. However, \verb$DF$ uses fewer global cables than \verb$SF$; thus, we expect that further commodization of optical cables will make the relative benefit of \verb$SF$ even bigger in the future.

\subsection{Energy Model}
\label{sec:power_results}

\goal{+Discuss the energy model and the results}

Energy consumption of interconnects can constitute 50\% of
the overall energy usage of a
computing center~\cite{Abts:2010:EPD:1815961.1816004}.
We now show that \verb$SF$ also offers substantial advantages in terms
of such operational costs. 
Following~\cite{Abts:2010:EPD:1815961.1816004} we assume that each router port has 4 lanes and there is one SerDes per lane consuming $\approx 0.7$ watts.
We compare
\verb$SF$ to other topologies using identical parameters as in the cost model.
We present the results in Figure~\ref{fig:power} and in Table~\ref{tab:detailed_cost_study}.
In general, \verb$SF$ is over 25\% more energy-efficient than \verb$DF$, \verb$FBF-3$, and \verb$DLN$. 
The power consumption in \verb$SF$ is lower than in
other topologies thanks to the lower number of routers and thus SerDes. 




\section{Discussion}

\goal{Introduce the section}

We demonstrated the Slim Fly topology which allows the construction of
low-latency, full-bandwidth, and resilient networks at a lower cost
than existing topologies. 

\subsection{Using Existing Routers}
\label{problemRadix}

\goal{+State the problem of limited designs and say we fix it}

Network architects often need to adjust to existing routers with a
given radix. 
As the construction of \verb$SF$ is based on powers of primes
$q$, network radices $k'$ (and thus router radices $k$) cannot have any arbitrary values for a simple
construction. We now illustrate solutions to this issue.

\goal{+Discuss the first solution}

First, the number of balanced \verb$SF$ constructions is significant. For network sizes up to 20,{}000, there are 11 balanced \verb$SF$ variants with full global bandwidth; \verb$DF$ offers only 8 such designs. Many of these variants can be directly constructed using readily available Mellanox routers with 18, 36, or 108 ports. Furthermore, the possibility of
applying oversubscription of $p$ with negligible effect on network
overall latency (see Section~\ref{sec:oversub}) adds even more flexibility to the construction of
network architectures based on \verb$SF$.

\goal{+Discuss how random channels fix the above problem}

Another option is to add random channels to utilize empty ports of
routers with radix $> k$ (using strategies presented in~\cite{Singla:2012:JND:2228298.2228322,DBLP:conf/isca/KoibuchiMAHC12}). 
This would additionally improve the latency and bandwidth of such \verb$SF$ variants~\cite{Singla:2012:JND:2228298.2228322,DBLP:conf/isca/KoibuchiMAHC12}. For
example, to construct a \verb$SF$ ($k=43$, $N=10830$)
with 48-port routers (cf., Aries~\cite{DBLP:conf/sc/FaanesBRCFAJKHR12}),
one could attach either five more endpoints or five random cables per
router. In order to minimize costs, one could also limit the random
connections to intra-rack copper links. We leave this analysis for future research.


\subsection{Constructing Dragonfly-type Networks}

\goal{+Discuss constructing df-type networks}

An interesting option is to use \verb$SF$ to implement
groups (higher-radix logical routers) of a \verb$DF$ or to connect
multiple groups of a \verb$DF$ topology. This could decrease the costs
in comparison to the currently used \verb$DF$
topologies~\cite{DBLP:conf/sc/FaanesBRCFAJKHR12,dally08}.

\subsection{Adding New Endpoints Incrementally}

\goal{+Show how SF handles incremental changes}

\verb$SF$ can seamlessly handle incremental changes in the number
of endpoints in computing centers. 
As we illustrated in the evaluation, the performance of \verb$SF$
is oblivious to relatively small oversubscription of $p$ and can still perform well
when $p > \lceil k'/2 \rceil$. It leaves 
a lot of flexibility for adding new endpoints incrementally. For example, 
a network with 10,830 endpoints can be extended by $\approx$1500 endpoints
before the performance drops by more than 10\%. To achieve this,
some ports in routers can be left empty and new endpoints would be added
with time according to the needs. This strategy is used in today's
Cray computing systems~\cite{DBLP:conf/sc/FaanesBRCFAJKHR12}.

\section{Related Work}

\goal{Discuss Related Work}

Related topologies are summarized in Section~\ref{sec:analysis}.  The main
benefits over traditional networks such as fat
tree~\cite{Leiserson:1985:FUN:4492.4495}, and
tori~\cite{Dally90performanceanalysis} are the significantly lower cost, energy
consumption, and latency. The advantages over state-of-the-art topologies such
as Flattened Butterfly~\cite{dally07} and Dragonfly~\cite{dally08} are higher
bandwidth, lower latency, in most cases higher resiliency, and lower (by $\approx$25-30\%)
cost and energy consumption.
The performance improvements are particularly important for data intense irregular
workloads~\cite{tate2014programming}, such as graph processing~\cite{besta2019substream, besta2019graph, gianinazzi2018communication, solomonik2017scaling, besta2017push, besta2017slimsell, besta2015accelerating, besta2015active, besta2018log, besta2019demystifying, besta2018survey} or others~\cite{schmid2016high, besta2014fault, gerstenberger2014enabling, kepner2016mathematical, besta2019slim, di2019network}, deep learning~\cite{ben2019modular}, or irregular matrix computations~\cite{kwasniewski2019red, besta2019communication}.
In fact, Slim Fly networks are related to those topologies in that they
minimize the diameter and reduce the number of routers while requiring longer
fiber cables. In comparison to random networks, \verb$SF$ does not rely on a
random construction for low diameter but starts from the lowest possible
diameter. As discussed in Section~\ref{problemRadix}, the ideas of random
shortcut topologies can be combined with Slim Flies.

Jiang et al.~\cite{Jiang:2009:IAR:1555754.1555783} propose indirect
adaptive routing algorithms for Dragonfly networks to balance the
traffic over the global links. Since the \topologyName topology is
homogeneous, it does not have isolated ``global links'' that could be overloaded
and backpressure is quickly propagated due to the low diameter. One can
use similar ideas to discover congestion in the second hop to make
better routing decisions for Slim Fly.

\section{Conclusion}

\goal{Repeat that we tackle an important challenge}

Interconnection networks constitute a significant part of the overall datacenter and HPC center construction and maintenance cost~\cite{Abts:2010:EPD:1815961.1816004}. Thus, reducing the cost and energy consumption of interconnects is an increasingly important task for the networking community. 

\goal{Repeat what SF is about}

We propose a new class of topologies called Slim Fly networks to
implement large datacenter and HPC network architectures. For this, we utilize a notion that lowering the network diameter reduces the amount of expensive network resources (cables, routers) used by packets traversing the network while maintaining high bandwidth. We define it as an \emph{optimization} problem and we optimize towards the Moore Bound. We then propose several techniques for designing optimal networks. We adopt a family of MMS graphs, which approach the Moore Bound for $D=2$, and we design Slim Fly basing on them.

\goal{Say SF is cheap and point to the future advances}

The Slim Fly architecture follows the technology trends towards high-radix routers and
cost-effective fiber optics. Under the current technology constraints, we achieve a 25\% cost and power
benefit over Dragonfly. We expect that further commodization of
fiber optics will lead to more cost-effective connections and
further improvements in silicon process technology will lead to higher-radix
routers. Both will make the relative benefit of Slim Fly even
bigger in the future.

\goal{Point out other advantages (routing, easy deploying)}

Our proposed routing strategies work well under bit permutation and
worst-case traffic patterns and asymptotically achieve high bandwidth
for random traffic. Thanks to the modular structure similar to Dragonfly, Slim Fly can be more easily deployed than other topologies such as random networks.

\goal{Discuss resiliency}

Theoretical analyses show that Slim Fly is more resilient to
link failures than Dragonfly and approaches highly resilient 
constructions such as random topologies. This counter-intuitive result (since the
topology utilizes less links and achieves a smaller diameter) can be
explained by the structure of the graph which has the properties of an
expander graph~\cite{Pippenger:1992:FCN:140901.141867}. 

\goal{Provide final, general conclusions}

Finally, the introduced approach for optimizing networks using the Moore Bound can be extended for higher-diameter networks which, while providing slightly higher latency, could establish scalable structures allowing for millions of endpoints. We believe that our general approach, based on
formulating engineering problems in terms of mathematical optimization, can effectively tackle other challenges in networking.

\bibliographystyle{abbrv}
\bibliography{sf_sc_2014}

\begin{thebibliography}{10}

\bibitem{abts2011cray}
D.~Abts.
\newblock {Cray XT4 and Seastar 3-D Torus Interconnect}.
\newblock {\em Encyclopedia of Parallel Computing}, pages 470--477, 2011.

\bibitem{Abts:2010:EPD:1815961.1816004}
D.~Abts, M.~R. Marty, P.~M. Wells, P.~Klausler, and H.~Liu.
\newblock {Energy Proportional Datacenter Networks}.
\newblock In {\em {Proceedings of the 37th Annual International Symposium on
  Computer Architecture}}, ISCA '10, pages 338--347, New York, NY, USA, 2010.
  ACM.

\bibitem{Alverson:2010:GSI:1901617.1902283}
R.~Alverson, D.~Roweth, and L.~Kaplan.
\newblock {The Gemini System Interconnect}.
\newblock In {\em {Proceedings of the 2010 18th IEEE Symposium on High
  Performance Interconnects}}, HOTI '10, pages 83--87, Washington, DC, USA,
  2010. IEEE Computer Society.

\bibitem{mellanox_director}
R.~Barriuso and A.~Knies.
\newblock {\em {108-Port InfiniBand FDR SwitchX Switch Platform Hardware User
  Manual}}, 2014.

\bibitem{ben2019modular}
T.~Ben-Nun, M.~Besta, S.~Huber, A.~N. Ziogas, D.~Peter, and T.~Hoefler.
\newblock A modular benchmarking infrastructure for high-performance and
  reproducible deep learning.
\newblock {\em arXiv preprint arXiv:1901.10183}, 2019.

\bibitem{bermond82}
J.~Bermond, C.~Delorme, and G.~Farhi.
\newblock {Large graphs with given degree and diameter III}.
\newblock {\em Annals of Discrete Mathematics}, 13:23--32, 1982.

\bibitem{besta2019slim}
M.~Besta et~al.
\newblock Slim graph: Practical lossy graph compression for approximate graph
  processing, storage, and analytics.
\newblock 2019.

\bibitem{besta2019substream}
M.~Besta, M.~Fischer, T.~Ben-Nun, J.~De~Fine~Licht, and T.~Hoefler.
\newblock Substream-centric maximum matchings on fpga.
\newblock In {\em ACM/SIGDA FPGA}, pages 152--161, 2019.

\bibitem{besta2014fault}
M.~Besta and T.~Hoefler.
\newblock Fault tolerance for remote memory access programming models.
\newblock In {\em ACM HPDC}, pages 37--48, 2014.

\bibitem{besta2015accelerating}
M.~Besta and T.~Hoefler.
\newblock Accelerating irregular computations with hardware transactional
  memory and active messages.
\newblock In {\em ACM HPDC}, 2015.

\bibitem{besta2015active}
M.~Besta and T.~Hoefler.
\newblock Active access: A mechanism for high-performance distributed
  data-centric computations.
\newblock In {\em ACM ICS}, 2015.

\bibitem{besta2018survey}
M.~Besta and T.~Hoefler.
\newblock Survey and taxonomy of lossless graph compression and space-efficient
  graph representations.
\newblock {\em arXiv preprint arXiv:1806.01799}, 2018.

\bibitem{besta2019communication}
M.~Besta, R.~Kanakagiri, H.~Mustafa, M.~Karasikov, G.~R{\"a}tsch, T.~Hoefler,
  and E.~Solomonik.
\newblock Communication-efficient jaccard similarity for high-performance
  distributed genome comparisons.
\newblock {\em arXiv preprint arXiv:1911.04200}, 2019.

\bibitem{besta2017slimsell}
M.~Besta, F.~Marending, E.~Solomonik, and T.~Hoefler.
\newblock Slimsell: A vectorizable graph representation for breadth-first
  search.
\newblock In {\em IEEE IPDPS}, pages 32--41, 2017.

\bibitem{besta2019demystifying}
M.~Besta, E.~Peter, R.~Gerstenberger, M.~Fischer, M.~Podstawski, C.~Barthels,
  G.~Alonso, and T.~Hoefler.
\newblock Demystifying graph databases: Analysis and taxonomy of data
  organization, system designs, and graph queries.
\newblock {\em arXiv preprint arXiv:1910.09017}, 2019.

\bibitem{besta2017push}
M.~Besta, M.~Podstawski, L.~Groner, E.~Solomonik, and T.~Hoefler.
\newblock To push or to pull: On reducing communication and synchronization in
  graph computations.
\newblock In {\em ACM HPDC}, 2017.

\bibitem{besta2019graph}
M.~Besta, D.~Stanojevic, J.~D.~F. Licht, T.~Ben-Nun, and T.~Hoefler.
\newblock Graph processing on fpgas: Taxonomy, survey, challenges.
\newblock {\em arXiv preprint arXiv:1903.06697}, 2019.

\bibitem{besta2018log}
M.~Besta, D.~Stanojevic, T.~Zivic, J.~Singh, M.~Hoerold, and T.~Hoefler.
\newblock Log (graph): a near-optimal high-performance graph representation.
\newblock In {\em PACT}, pages 7--1, 2018.

\bibitem{Bollobas01:random_graphs}
B.~Bollobas.
\newblock {\em {Random Graphs}}.
\newblock Cambridge University Press, 2001.

\bibitem{Chen:2012:LUH:2388996.2389090}
D.~Chen, N.~Eisley, P.~Heidelberger, S.~Kumar, A.~Mamidala, F.~Petrini,
  R.~Senger, Y.~Sugawara, R.~Walkup, B.~Steinmacher-Burow, A.~Choudhury,
  Y.~Sabharwal, S.~Singhal, and J.~J. Parker.
\newblock {Looking Under the Hood of the IBM Blue Gene/Q Network}.
\newblock In {\em Proceedings of the ACM/IEEE Supercomputing}, SC '12, pages
  69:1--69:12, Los Alamitos, CA, USA, 2012. IEEE Computer Society Press.

\bibitem{Chen:2011:IBG:2063384.2063419}
D.~Chen, N.~A. Eisley, P.~Heidelberger, R.~M. Senger, Y.~Sugawara, S.~Kumar,
  V.~Salapura, D.~L. Satterfield, B.~Steinmacher-Burow, and J.~J. Parker.
\newblock {The IBM Blue Gene/Q Interconnection Network and Message Unit}.
\newblock In {\em {Proceedings of 2011 ACM/IEEE Supercomputing}}, SC '11, pages
  26:1--26:10, New York, NY, USA, 2011. ACM.

\bibitem{Dally:2003:PPI:995703}
W.~Dally and B.~Towles.
\newblock {\em {Principles and Practices of Interconnection Networks}}.
\newblock Morgan Kaufmann Publishers Inc., San Francisco, CA, USA, 2003.

\bibitem{Dally90performanceanalysis}
W.~J. Dally.
\newblock {Performance Analysis of k-ary n-cube Interconnection Networks}.
\newblock {\em IEEE Transactions on Computers}, 39:775--785, 1990.

\bibitem{delorme85}
C.~Delorme.
\newblock {Grands Graphes de Degr\'{e}e et Diam\`{e}tre Donn\'{e}s}.
\newblock {\em Europ. J. Combinatorics}, 6:291--302, 1985.

\bibitem{di2019network}
S.~Di~Girolamo, K.~Taranov, A.~Kurth, M.~Schaffner, T.~Schneider,
  J.~Ber{\'a}nek, M.~Besta, L.~Benini, D.~Roweth, and T.~Hoefler.
\newblock Network-accelerated non-contiguous memory transfers.
\newblock {\em arXiv preprint arXiv:1908.08590}, 2019.

\bibitem{domke-hoefler-dfsssp}
J.~Domke, T.~Hoefler, and W.~Nagel.
\newblock {Deadlock-Free Oblivious Routing for Arbitrary Topologies}.
\newblock In {\em {Proceedings of the 25th IEEE International Parallel and
  Distributed Processing Symposium (IPDPS)}}, pages 613--624. IEEE Computer
  Society, May 2011.

\bibitem{dongarra2013visit}
J.~Dongarra.
\newblock {Visit to the National University for Defense Technology Changsha,
  China}.
\newblock {\em Oak Ridge National Laboratory, Tech. Rep., June}, 2013.

\bibitem{Duato:2002:INE:572400}
J.~Duato, S.~Yalamanchili, and N.~Lionel.
\newblock {\em Interconnection Networks: An Engineering Approach}.
\newblock Morgan Kaufmann Publishers Inc., San Francisco, CA, USA, 2002.

\bibitem{DBLP:conf/sc/FaanesBRCFAJKHR12}
G.~Faanes, A.~Bataineh, D.~Roweth, T.~Court, E.~Froese, R.~Alverson,
  T.~Johnson, J.~Kopnick, M.~Higgins, and J.~Reinhard.
\newblock {Cray cascade: a scalable HPC system based on a Dragonfly network}.
\newblock In {\em {SC}}, page 103. IEEE/ACM, 2012.

\bibitem{Flich:2012:SET:2360762.2361136}
J.~Flich, T.~Skeie, A.~Mejia, O.~Lysne, P.~Lopez, A.~Robles, J.~Duato,
  M.~Koibuchi, T.~Rokicki, and J.~C. Sancho.
\newblock {A Survey and Evaluation of Topology-Agnostic Deterministic Routing
  Algorithms}.
\newblock {\em IEEE Trans. Parallel Distrib. Syst.}, 23(3):405--425, Mar. 2012.

\bibitem{gerstenberger2014enabling}
R.~Gerstenberger, M.~Besta, and T.~Hoefler.
\newblock Enabling highly-scalable remote memory access programming with mpi-3
  one sided.
\newblock {\em Scientific Programming}, 22(2):75--91, 2014.

\bibitem{gianinazzi2018communication}
L.~Gianinazzi, P.~Kalvoda, A.~De~Palma, M.~Besta, and T.~Hoefler.
\newblock Communication-avoiding parallel minimum cuts and connected
  components.
\newblock In {\em ACM SIGPLAN Notices}, volume~53, pages 219--232. ACM, 2018.

\bibitem{4228210}
C.~Gomez, F.~Gilabert, M.~Gomez, P.~Lopez, and J.~Duato.
\newblock Deterministic versus adaptive routing in fat-trees.
\newblock In {\em Parallel and Distributed Processing Symposium, 2007. IPDPS
  2007. IEEE International}, pages 1--8, March 2007.

\bibitem{Gopal:1994:PSD:201173.201226}
I.~S. Gopal.
\newblock {Interconnection Networks for High-performance Parallel Computers}.
\newblock chapter {Prevention of Store-and-forward Deadlock in Computer
  Networks}, pages 338--344. IEEE Computer Society Press, Los Alamitos, CA,
  USA, 1994.

\bibitem{Hafner2004223}
P.~R. Hafner.
\newblock {Geometric realisation of the graphs of McKay–Miller–Širáň}.
\newblock {\em Journal of Combinatorial Theory, Series B}, 90(2):223 -- 232,
  2004.

\bibitem{Jiang:2009:IAR:1555754.1555783}
N.~Jiang, J.~Kim, and W.~J. Dally.
\newblock {Indirect Adaptive Routing on Large Scale Interconnection Networks}.
\newblock In {\em {Proceedings of the 36th Annual International Symposium on
  Computer Architecture}}, ISCA '09, pages 220--231, New York, NY, USA, 2009.
  ACM.

\bibitem{karypis99}
G.~Karypis and V.~Kumar.
\newblock {A Fast and Highly Quality Multilevel Scheme for Partitioning
  Irregular Graphs}.
\newblock {\em SIAM Journal on Scientific Computing}, 20:359--392, 1999.

\bibitem{kepner2016mathematical}
J.~Kepner, P.~Aaltonen, D.~Bader, A.~Bulu{\c{c}}, F.~Franchetti, J.~Gilbert,
  D.~Hutchison, M.~Kumar, A.~Lumsdaine, H.~Meyerhenke, et~al.
\newblock Mathematical foundations of the graphblas.
\newblock In {\em 2016 IEEE High Performance Extreme Computing Conference
  (HPEC)}, pages 1--9. IEEE, 2016.

\bibitem{Kim:2007:FBT:1331699.1331717}
J.~Kim, J.~Balfour, and W.~Dally.
\newblock {Flattened Butterfly Topology for On-Chip Networks}.
\newblock In {\em Proceedings of the 40th Annual IEEE/ACM International
  Symposium on Microarchitecture}, MICRO 40, pages 172--182, Washington, DC,
  USA, 2007. IEEE Computer Society.

\bibitem{dally07}
J.~Kim, W.~J. Dally, and D.~Abts.
\newblock {Flattened Butterfly: A Cost-efficient Topology for High-radix
  Networks}.
\newblock In {\em Proceedings of the 34th Annual International Symposium on
  Computer Architecture}, ISCA '07, pages 126--137, New York, NY, USA, 2007.
  ACM.

\bibitem{dally08}
J.~Kim, W.~J. Dally, S.~Scott, and D.~Abts.
\newblock {Technology-Driven, Highly-Scalable Dragonfly Topology}.
\newblock In {\em Proceedings of the 35th Annual International Symposium on
  Computer Architecture}, ISCA '08, pages 77--88, Washington, DC, USA, 2008.
  IEEE Computer Society.

\bibitem{DBLP:conf/isca/KoibuchiMAHC12}
M.~Koibuchi, H.~Matsutani, H.~Amano, D.~F. Hsu, and H.~Casanova.
\newblock {A case for random shortcut topologies for HPC interconnects}.
\newblock In {\em {ISCA'12}}, pages 177--188. IEEE, 2012.

\bibitem{kwasniewski2019red}
G.~Kwasniewski, M.~Kabi{\'c}, M.~Besta, J.~VandeVondele, R.~Solc{\`a}, and
  T.~Hoefler.
\newblock Red-blue pebbling revisited: near optimal parallel matrix-matrix
  multiplication.
\newblock In {\em ACM/IEEE Supercomputing}, page~24. ACM, 2019.

\bibitem{Leiserson:1985:FUN:4492.4495}
C.~E. Leiserson.
\newblock {Fat-trees: universal networks for hardware-efficient
  supercomputing}.
\newblock {\em IEEE Trans. Comput.}, 34(10):892--901, Oct. 1985.

\bibitem{lidl1997finite}
R.~Lidl and H.~Niederreiter.
\newblock {Finite Fields: Encyclopedia of Mathematics and Its Applications.}
\newblock {\em Computers \& Mathematics with Applications}, 33(7):136--136,
  1997.

\bibitem{mckay98}
B.~D. McKay, M.~Miller, and J.~\v{S}ir\'{a}n.
\newblock A note on large graphs of diameter two and given maximum degree.
\newblock {\em Journal of Combinatorial Theory, Series B}, 74(1):110 -- 118,
  1998.

\bibitem{miller2005}
M.~Miller and J.~{\v{S}}ir{\'a}n.
\newblock {Moore graphs and beyond: A survey of the degree/diameter problem}.
\newblock {\em Electronic Journal of Combinatorics}, 61:1--63, 2005.

\bibitem{Pippenger:1992:FCN:140901.141867}
N.~Pippenger and G.~Lin.
\newblock {Fault-tolerant circuit-switching networks}.
\newblock In {\em {Proceedings of the Fourth Annual ACM Symposium on Parallel
  Algorithms and Architectures}}, SPAA '92, pages 229--235, New York, NY, USA,
  1992. ACM.

\bibitem{schmid2016high}
P.~Schmid, M.~Besta, and T.~Hoefler.
\newblock High-performance distributed rma locks.
\newblock In {\em ACM HPDC}, pages 19--30, 2016.

\bibitem{Scott:2006:BHC:1135775.1136488}
S.~Scott, D.~Abts, J.~Kim, and W.~J. Dally.
\newblock {The BlackWidow High-Radix Clos Network}.
\newblock In {\em {Proceedings of the 33rd annual International Symposium on
  Computer Architecture}}, ISCA '06, pages 16--28, Washington, DC, USA, 2006.
  IEEE Computer Society.

\bibitem{ugal2005scheme}
A.~Singh.
\newblock {\em {Load-Balanced Routing in Interconnection Networks}}.
\newblock PhD thesis, Stanford University, 2005.

\bibitem{Singla:2012:JND:2228298.2228322}
A.~Singla, C.-Y. Hong, L.~Popa, and P.~B. Godfrey.
\newblock {Jellyfish: networking data centers randomly}.
\newblock In {\em {Proceedings of the 9th USENIX conference on Networked
  Systems Design and Implementation}}, NSDI'12, pages 17--17, Berkeley, CA,
  USA, 2012. USENIX Association.

\bibitem{solomonik2017scaling}
E.~Solomonik, M.~Besta, F.~Vella, and T.~Hoefler.
\newblock Scaling betweenness centrality using communication-efficient sparse
  matrix multiplication.
\newblock In {\em ACM/IEEE Supercomputing}, page~47, 2017.

\bibitem{tate2014programming}
A.~Tate et~al.
\newblock Programming abstractions for data locality.
\newblock PADAL Workshop 2014, 2014.

\bibitem{Tiyyagura:2008:TSP:1361718.1361719}
S.~Tiyyagura, P.~Adamidis, R.~Rabenseifner, P.~Lammers, S.~Borowski,
  F.~Lippold, F.~Svensson, O.~Marxen, S.~Haberhauer, A.~Seitsonen,
  J.~Furthm\"{u}ller, K.~Benkert, M.~Galle, T.~B\"{o}nisch, U.~K\"{u}ster, and
  M.~Resch.
\newblock {Teraflops Sustained Performance With Real World Applications}.
\newblock {\em Int. J. High Perform. Comput. Appl.}, 22(2):131--148, May 2008.

\bibitem{2013arXiv1301.4177T}
R.~V. {Tomic}.
\newblock {Network Throughput Optimization via Error Correcting Codes}.
\newblock {\em ArXiv e-prints}, Jan. 2013.

\bibitem{valiant1982scheme}
L.~Valiant.
\newblock {A scheme for fast parallel communication}.
\newblock {\em SIAM journal on computing}, 11(2):350--361, 1982.

\bibitem{siagiova01}
J.~\v{S}iagiov\'{a}.
\newblock {A Note on the McKay-Miller-\v{S}ir\'{a}n Graphs}.
\newblock {\em Journal of Combinatorial Theory, Series B}, 81:205--208, 2001.

\bibitem{pleiades}
R.~Wolf.
\newblock {Nasa Pleiades Infiniband Communications Network}, 2009.
\newblock Intl. ACM Symposium on High Performance Distributed Computing.

\bibitem{Yuan:2013:NRS:2503210.2503229}
X.~Yuan, S.~Mahapatra, W.~Nienaber, S.~Pakin, and M.~Lang.
\newblock {A New Routing Scheme for Jellyfish and Its Performance with HPC
  Workloads}.
\newblock In {\em Proceedings of 2013 ACM/IEEE Supercomputing}, SC '13, pages
  36:1--36:11, 2013.

\end{thebibliography}

\end{document}